\documentclass[%
 aip,
 amsmath,amssymb,
 reprint,%
]{revtex4-1}

\usepackage{array}[=2016-10-06]
\usepackage{graphicx}
\usepackage{dcolumn}
\usepackage{bm}

\usepackage[utf8]{inputenc}
\usepackage[T1]{fontenc}
\usepackage{mathptmx}
\usepackage{etoolbox}
\usepackage{dcolumn}
\makeatletter
\def\@email#1#2{%
 \endgroup
 \patchcmd{\titleblock@produce}
  {\frontmatter@RRAPformat}
  {\frontmatter@RRAPformat{\produce@RRAP{*#1\href{mailto:#2}{#2}}}\frontmatter@RRAPformat}
  {}{}
}%
\makeatother
\begin{document}

\preprint{AIP/123-QED}

\title[ALD optimization using reasoning LLMs]{Performance of AI agents based on reasoning language models on ALD process optimization tasks}
\author{Angel Yanguas-Gil$^*$}
\email{ayg@anl.gov}
\affiliation{ 
Applied Materials Division, Argonne National Laboratory, Lemont, IL 60439, USA
}%

\date{\today}

\begin{abstract}
In this work we explore the performance and behavior of reasoning large language models to autonomously optimize atomic layer deposition (ALD) processes.
In the ALD process optimization task, an agent built on top of a reasoning LLM has to find optimal dose times for an ALD precursor and a coreactant without any prior knowledge on the process, including whether it is actually self-limited. The agent is meant to interact iteratively with an ALD reactor in a fully unsupervised way. We evaluate this agent using a simple model of an ALD tool that incorporates ALD processes with different self-limited surface reaction pathways as well as a non self-limited component.
Our results show that agents based on reasoning models like OpenAI's o3 and GPT5 consistently succeeded at completing this optimization task. However, we observed significant run-to-run variability due to the non deterministic nature of the model's response. In order to understand the logic followed by the reasoning model, the agent uses a two step process in which the model first generates an open response detailing the reasoning process. This response is then transformed into a structured output. An analysis of these reasoning traces showed that the logic of the model was sound and that its reasoning was based on the notions of self-limited process and saturation expected in the case of ALD. However, the agent can sometimes be misled by its own prior choices when exploring the optimization space.
\end{abstract}

\maketitle

\section{Introduction}

The field of generative AI has been yielding models with increasing levels of performance. One of the most significant recent breakthroughs has
been the development of the so-called reasoning large language models. These models, which include examples such as OpenAI's o1 and o3, Qwen/QwQ, and DeepSeek's open-weights R1\cite{Yang2025,Guo_2025}, are significantly better than traditional LLMs at so called reasoning tasks in areas such as math and coding\cite{balunovic2025,zhang2025}. Many of the commercially available models as of early 2026 are hybrid models, directing complex queries to these more advanced models.

How to effectively leverage these models in the physical sciences is still an active area of research. One application is to act as a core component of autonomous materials synthesis platforms. However, our understanding of LLMs' capabilities in the context of thin film growth is still lacking. This is a problem shared across the physical sciences, where the development of general methodologies to evaluate these models in useful, practical contexts is still in the early stages of
development.

In this work we explore the potential of reasoning large language models in the context of atomic layer deposition. In particular, we focus on the optimization of a traditional AB-type ALD process comprising a precursor and a coreactant. This is a common task both for the characterization of processes based on new ALD precursors and when adapting existing processes to a specific ALD reactor. It is also a process that has a clear outcome: an optimized ALD process is one where the dose times are chosen so that the process is self-limited and the growth per cycle is very close to its saturated value while ideally minimizing the total deposition time. While some aspects are open to subjective analysis, domain experts can evaluate saturation curves and
understand the implications of choosing a specific dose time. Moreover,
prior works have explored this problem from a machine learning perspective, offering a way to compare the performance of reasoning large language models with respect to more classical approaches\cite{Paulson2021}. There is also a wealth of experimental data in the literature on the description and optimization of new ALD processes, a task that is generally carried out by obtaining saturation curves for both the precursor and co-reactant\cite{Aaltonen2005,Comstock2012,Devika2020,Hamalainen2012,Klaus2000,Klepper2007,Tero2009}. This gives us a good sense of how many samples and the steps that a human expert would follow when carrying out this task and therefore a benchmark for a model performance.

We therefore seek to answer the following two questions: 1) how good are reasoning models at process optimization tasks? 2) how do reasoning models actually reason about ALD process optimization? The first question is multifaceted, and it includes gauging the model's ability to identify the optimal conditions, analyzing its run-to-run variability, quantifying the number of samples required and the impact of initial conditions and noise. The second question requires having access to the 
model's reasoning process and ensure that the suggested additional experiments are consistent with this logic.

To answer these questions, we have adapted a functional model of an ALD reactor used in a prior work to explore process optimization using
machine learning\cite{Paulson2021}. We have expanded this model to consider additional situations, including the presence of a non self-limited component. Details of
this model will be presented in Section \ref{sec:aldmodel}. This
gives us full control of the underlying ALD process, including the quality of the feedback provided to the model and the nature of the ALD process, from ideal self-limited to soft-saturating and processes with non self-limited component.
We also introduce an agent that iteratively interacts with the ALD reactor in a two step process: in the first step, the model is asked to respond with suggestions for new growth conditions using an open ended response format. This allows us to capture the reasoning steps the model has followed in each iteration. In a second step, this open ended response format is subsequently transformed into a structured output that includes a determination on whether the process has been optimized and the suggested next set of conditions. 

\section{Methodology}

\subsection{Reasoning large language models}

Reasoning large language models are an evolution of traditional large language models that overcome some of LLMs' limitations. While a traditional LLM is a feedforward model that stochastically generates outputs based on inputs, a reasoning large language model has a more complex structure designed for problem decomposition and validation prior to the generation of the final outputs. One of the techniques used as building blocks for these models is the so-called chain-of-thought process that tries to break down a problem into multiple independent steps\cite{Wei2023}. Conventional LLMs as well as other algorithms are used to address these specific steps and generate a final response. As part of this process, reasoning language models generate intermediate outputs that are used by these additional components to solve complex problems beyond the scope of traditional LLMs.

In a prior work, we explored LLMs' ability to answer queries focused on atomic layer deposition\cite{YanguasGil2025}. The most advanced model used in that work was GPT4o, which is a conventional LLM. In this work we use a reasoning large language model as a building block to construct an agent capable of autonomously carrying out the optimization of an ALD process. In particular, we focus on two reasoning LLMs: o3, a pure reasoning model from OpenAI, and GPT5, a hybrid model where complex queries are processed using models with reasoning capabilities. However, the methodologies used in this work are general and model agnostic and could be used with any other commercial and open weight model. Other examples of reasoning models include Qwen/QwQ and DeepSeek R1\cite{Yang2025,Guo_2025}. 

\subsection{ALD process optimization agent}

\begin{figure}
\includegraphics[width=8cm]{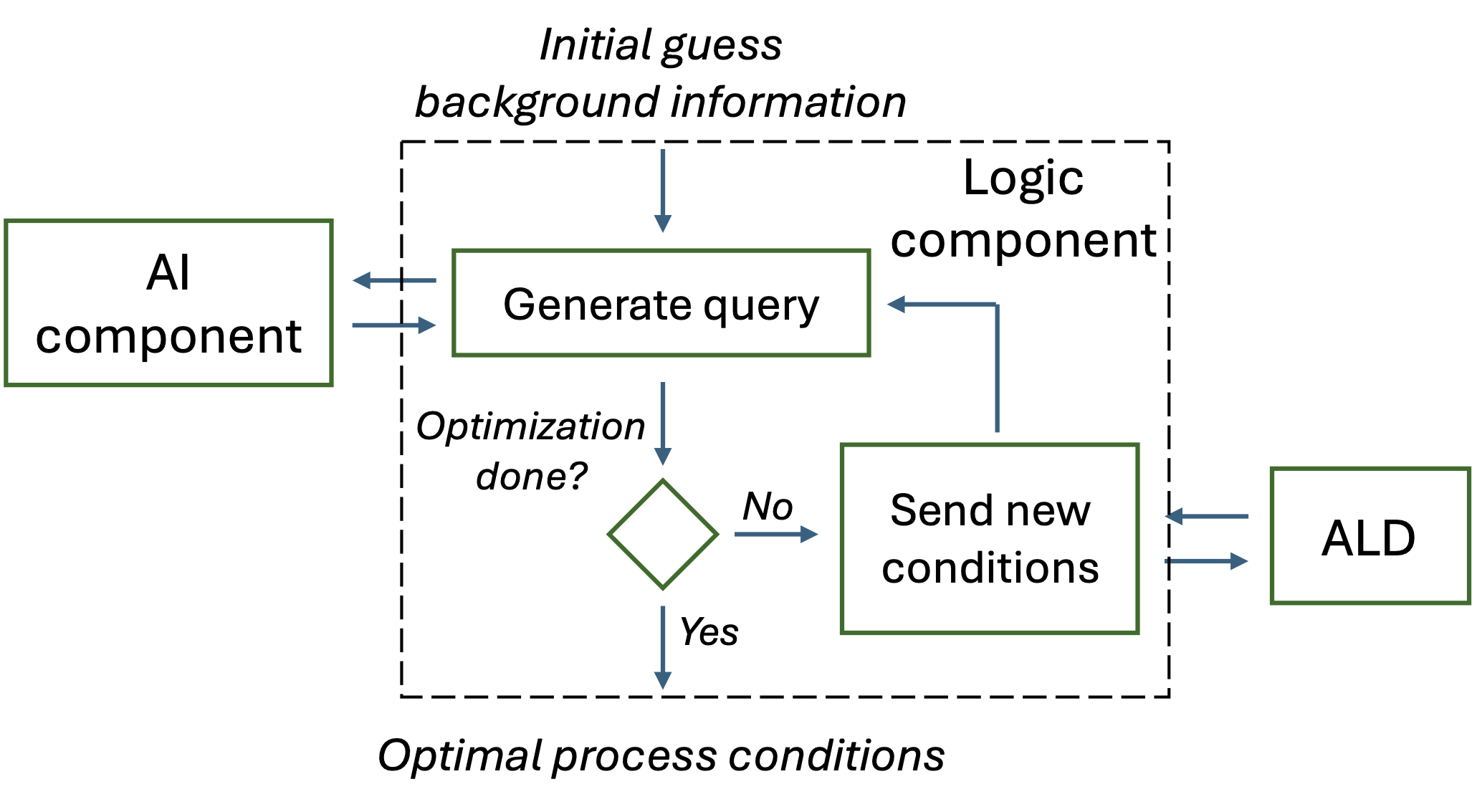}
\caption{\label{fig:agent}  Scheme of our AI agent for ALD process optimization: the logic component generates queries and process the response of the AI component, which uses a reasoning model to determine the strategy for optimization and request additional experiments. Based on the AI component's response, the logic component sends new conditions to the simulated ALD reactor.}
\end{figure}

Our optimization process relies on a simple agent architecture meant to interact with an ALD reactor (in this case a simulated reactor). This is compatible with our
experimental ALD tool designed to integrate with generative AI. The agent architecture, shown in Figure \ref{fig:agent}, has a logic and an AI component. The logic component iteratively asks the AI component to identify the optimal dose times for the precursor and coreactant that leads to a saturated GPC or to request new experiments to be carried out by the ALD reactor. Based on this response, the logic component
requests additional growths to the simulated ALD reactor or terminates the optimization process, returning the optimal conditions.

Since one of the goals of this work is to gain insights on the strategies and chain of thought used by reasoning models during the optimization of ALD processes,
the AI component carries out two consecutive calls to the reasoning LLM during each iteration: the first call provides the context and instruction to
the model, including any existing experimental data (in this case growth per cycle for different precursor and coreactant dose times) and any prior information about the process (e.g. whether the precursor has a high or a low vapor pressure, or the expected growth per cycle for a known process). The model provides an open response on how to proceed, including suggestions for new conditions, and the determination of whether the process has been optimized and whether it is actually self-limited. A second call takes this open response and transforms it into a structured output containing a list of experimental conditions and two flags identifying if the process is optimized or not self-limited. The model is asked to generate this response in JavaScript Object Notation (JSON) format. Consequently, for each iteration we have the open response with the model reasoning and a clear output that the procedural component can work with.

We consider two different agent variants: in our base agent, the reasoning model receives the prompt and the past information about the growth during each iteration when asked how to proceed. In the memory variant, the reasoning language model also receives the model output for all prior iterations. This provides some continuity to the reasoning process across iterations.

\begin{figure*}
\includegraphics[width=17cm]{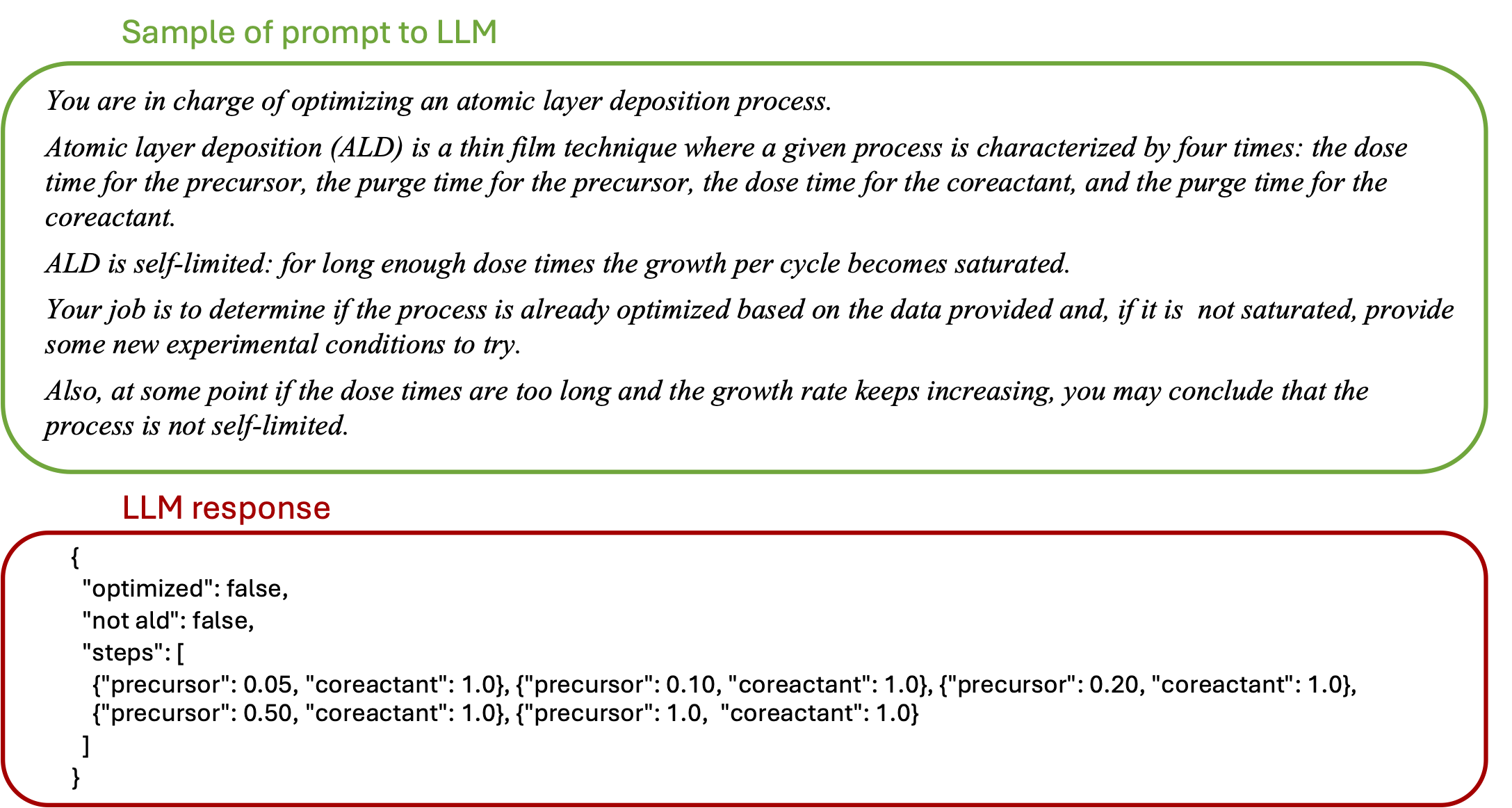}
\caption{\label{fig:scheme}  Sample of the prompt passed
to the reasoning model as well as a sample of the structured
JSON returned by the model during each iteration. The model
response determines whether the agent concludes the optimization
or continues requesting additional experiments}
\end{figure*}

Figure \ref{fig:scheme} shows an example of the original prompt passed to the reasoning model. Beyond a basic explanation of self-limited processes, there are no specific instructions about how to carry out the optimization process. Some variations of this prompt will be explored in Section \ref{sec:noguess}. A full description of the prompts is provided in Appendix \ref{sec:prompt}.

\subsection{\label{sec:aldmodel}ALD process model}

In this work we interface the AI agent with a simulated ALD process. Our model builds on a model developed in prior work\cite{YanguasGil2014,Yanguasgil2021} and has been used in the past to evaluate machine learning algorithms for ALD process optimization\cite{Paulson2021}. We consider a self-limited surface kinetics comprising one or more surface reaction pathways, each characterized by a first order irreversible Langmuir kinetics for the precursor and the coreactant. For each of these pathways the state of the surface can be expressed in terms of a fractional surface coverage $\theta_i$ that describes the fraction of surface sites that are occupied by dissociatively adsorbed precursor molecules. The total fractional surface coverage is the sum of each of the individual components multiplied by a scaling factor $f_i$ that determines the relative prevalence of each reaction pathway:
\begin{equation}
\theta = \sum_i f_i \theta_i
\end{equation}
with $\sum_i f_i = 1$. With this approximation, in a fully self-limited process where the precursor and co-reactant are perfectly isolated by purges, the growth per cycle is equal to the saturated GPC multiplied by the total surface coverage
change during the precursor dose:
\begin{equation}
\mathrm{GPC} = \mathrm{GPC}_0 (\theta-\theta_0)
\end{equation}
where $\theta_0$ is the surface coverage of reacted precursor at the beginning of the precursor dose and $\theta$ the surface coverage at the end of the precursor dose.
Each reaction pathway is also determined by two reaction constants $k_1$ and $k_2$ that are the inverse of the characteristic time for saturation. These constants incorporate factors such as the precursor pressure and the sticking probability. It can be shown (see Appendix \ref{sec:appendixmodel}) that the steady state GPC of a self-limited process
depends on the precursor and coreactant dose times $t_1$ and $t_2$ as follows:
\begin{equation}
\mathrm{GPC} = \mathrm{GPC}_0\frac{(1-e^{-k_1t_1})(1-e^{-k_2t_2})}{1-e^{-(k_1t_1+k_2t_2)}}
\end{equation}
For a process involving multiple reaction pathways, each component $\theta_i$ has its own two characteristic constants $k^i_1$ and $k^i_2$, so that the GPC is
simply:
\begin{equation}
\mathrm{GPC} = \mathrm{GPC}_0 \sum_i f_i \frac{(1-e^{-k^i_1t_1})(1-e^{-k^i_2t_2})}{1-e^{-(k^i_1t_1+k^i_2t_2)}}
\end{equation}
 
We can extend the model to consider the presence of a CVD component. One effective way of introducing this component is to consider that, during the precursor dose, there is a non-zero background pressure of coreactant that gives a constant growth rate superimposed to the self-limited precursor-surface interaction.
This requires adding a third constant $k_c$ to the kinetic model to represent the co-reactant interaction with the surface during the precursor dose.
Given a CVD growth rate $\mathrm{GR}_0$, the
growth can be calculated analytically from the precursor and coreactant dose  times $t_1$ and $t_2$ using the following expression:
\begin{equation}
\mathrm{GPC}  =  \mathrm{GR}_0 t_1 + \mathrm{GPC}_0\theta_\mathrm{lim} (\theta_\mathrm{lim}-\theta_0)(1-e^{-(k_1+k_c)t_1})
\end{equation}
where
\begin{eqnarray}
\theta_0  & = & \frac{e^{-k_2t_2}(1-e^{-(k_1+k_c)t_1})}{1-e^{-(k_1+k_c)t_1}e^{-k_2t_2}} \\
\theta_\mathrm{lim} &=& \frac{k_1}{k_c+k_1} \\
k_c & =& \frac{k_1\mathrm{GR}_0}{k_1\mathrm{GPC}_0-\mathrm{GR}_0}
\end{eqnarray}
Again, details can be found in Appendix \ref{sec:appendixmodel}.

\begin{figure}
\includegraphics[width=8cm]{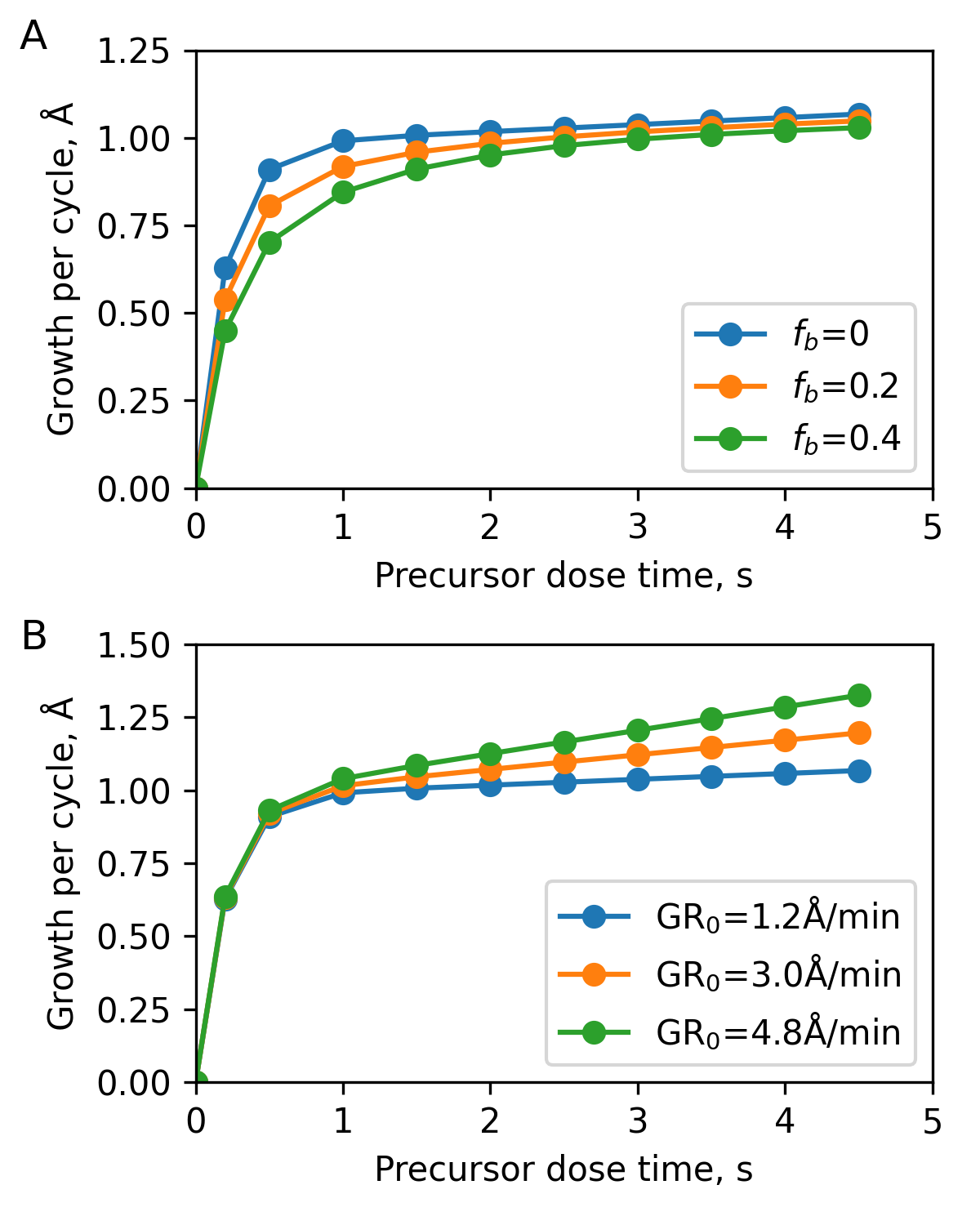}
\caption{\label{fig:sat}Precursor saturation curves of simulated ALD processes used to evaluate the process optimization by AI agents: A) Soft-saturating model, showing the
impact of a second, lower reactivity reaction pathway ($k^a_1$ = 5 s$^{-1}$, $k^b_1$ = 1 s$^{-1}$, $k_2$ = 4 s$^{-1}$); B) Impact of CVD component for various values of the non self-limited CVD component ($k_1$ = 5 s$^{-1}$, $k_2$ = 4 s$^{-1}$). In both cases, the coreactant dose time was set to $t_2$ = 1~s.}
\end{figure}

While more sophisticated surface kinetic models have been reported in the literature, this simple parametric model captures many of the phenomenological behavior
of ALD processes. For instance, we can incorporate soft-saturating ALD processes with a fast and a slow reaction component: in Fig. \ref{fig:sat}(A) we show the impact of having a second reaction pathway characterized by a saturation time that is 5 times
slower as a function of the relative weight of the
slower reaction pathway, $f_2$.  In Figure \ref{fig:sat}(B) we show saturation curves with different degrees of non self-limited components.

\begin{table}
\caption{\label{tab:processes}Main ALD processes used in this work.}
\begin{ruledtabular}
\begin{tabular}{l|l}
 Name & Model parameters  \\
\hline
fast/fast, 1\AA/cycle & $k_1$ = 5 s$^{-1}$, $k_2$ = 4 s$^{-1}$,  GPC$_0$ = 1\AA  \\
slow/slow, 1\AA/cycle & $k_1$ = 1 s$^{-1}$, $k_2$ = 1.2 s$^{-1}$,  GPC$_0$ = 1\AA  \\
slow/fast, 1\AA/cycle & $k_1$ = 1 s$^{-1}$, $k_2$ = 4 s$^{-1}$,  GPC$_0$ = 1\AA  \\
fast/fast, 0.3\AA/cycle & $k_1$ = 5 s$^{-1}$, $k_2$ = 4 s$^{-1}$,  GPC$_0$ = 0.3\AA  \\
soft/fast,  1\AA/cycle & $k^a_1$ = 5 s$^{-1}$, $k^b_1$ = 1 s$^{-1}$, $f_b$ = 0.2, \\
& $k_2$ = 4 s$^{-1}$,  GPC$_0$ = 1\AA  \\
\end{tabular}
\end{ruledtabular}
\end{table}

We have chosen a few prototypical ALD processes for this work. These are shown in Table \ref{tab:processes}. They intend to cover a representative sample of commonly observed behaviors: from a saturation curve perspective, we want to cover four different regimes, which we refer to as fast/fast, fast/slow, slow/slow, and soft/fast processes. A fast half cycle is defined as one where saturation is reached for dose times of the order of 1~s. A slow process is a process requiring dose times of at least 5~s or longer. We 
also considered two values of saturated growth per cycles, 1.0 and 0.3~\AA/cycle. 
In addition to the ideal models we will also consider the impact of noise in the feedback from the reactor as well as the presence of non self-limited components.

\section{Results}

\subsection{\label{sec:performance}Performance of base agent}

We have evaluated the base agent's ability to optimize each of the five ALD processes listed in Table \ref{tab:processes}. We have tracked agent self-reported success on optimizing the ALD process, the selected precursor and coreactant dose times, the corresponding growth per cycle, the number of experiments (hereafter referred to as samples) required to carry out the optimization, and the number of iterations required for the agent to complete (or give up) the optimization process. We have
first considered the case where the agent is provided a first guess on the optimal dose times. We have evaluated two initial conditions: 0.2~s and 2~s for both the precursor and coreactant. These are referred to as (0.2~s,0.2~s) and (2~s, 2~s). 
When optimizing either of these processes, the agent does not have any prior information on the expected growth per cycle, 
which represents the optimization of a new ALD process. Therefore, we are evaluating the reasoning language models in the worst-case scenario of not having any prior information about either the current ALD process or any past ALD process optimized for this reactor to base their decision making process on.
For each condition we have carried out ten independent runs. The results are summarized in Table \ref{tab:megatable}.

\begin{table*}
\caption{\label{tab:megatable}Performance of the AI agent during the optimization of different ALD processes. In all cases, 
averages over
ten independent runs are provided, with the number in parenthesis representing the standard deviation.}
\begin{ruledtabular}
\begin{tabular}{ccc|ccccccc}
ALD Process & Guess & Model & Success (\%) & $t_1$, s & $t_2$, s & GPC, \AA/cy & \# Samples & \# Iter & $t_1+t_2$, s \\
\hline
fast/fast& 0.2 s, 0.2 s &  o3& 100  & 0.86(0.15)        & 1.3(0.3)       & 0.97(0.02)  & 13(5)     & 5(1)   & 2.2(0.4) \\
  1.0\AA/cycle & & GPT5 & 100 & 0.87(0.11)       & 1.1(0.2)        & 0.97(0.02)  & 11(3)    & 4.2(1.0)   & 2.0(0.3) \\
 & 2 s, 2 s     & o3 &  100   & 1.6(0.4)        & 1.6(0.4)        &  0.99(0.01)        & 9(3)     & 2.6(0.7)    & 3.2(0.9) \\
 & & GPT5     & 100 & 1.8(0.4)        & 1.8(0.3)   & 0.99(0.01)          & 4.7(1.1)     & 2.5(0.5)   & 3.6(0.7) \\
\hline
slow/slow  & 0.2 s, 0.2 s  & o3  & 100 & 3.2(1.0) & 3.6(0.7) & 0.92(0.08) & 26(6) & 8.0(1.7) & 6.8(1.5) \\
 1.0\AA/cycle &   & GPT5 & 100 & 2.5(0.9) & 3.2(0.5) & 0.86(0.11) & 18(5) & 6(2)     & 5.6(1.2) \\
 &  2 s, 2 s    &  o3    & 100 & 5.3(0.9) & 4.8(0.9) & 0.99(0.01) & 10(2) & 3.9(0.8) & 10.1(1.4) \\
 &     & GPT5   & 100 & 5.5(0.8) & 4.7(0.6) & 0.99(0.01) & 8(2)  & 3.8(1.0) & 10.2(1.2) \\
 \hline
slow/fast & 0.2 s, 0.2 s & o3  & 100   & 3.8(0.6)  & 1.4(0.3) & 0.97(0.01) &  23(5)   & 6.7(1.2) & 5.2(0.6) \\
1.0\AA/cycle & & GPT5 & 100 & 3.4(0.8)   & 1.2(0.2)    & 0.95(0.04)  &   16(5)  & 7(2)  & 4.6(0.9) \\
   & 2 s, 2 s  & o3  & 100   &     3.8(0.6)    &  1.0(0.4)    &  0.97(0.01)   & 9(2)   &   3.3(0.4)      & 5.8(0.9) \\
& & GPT5     & 100   &  5.2(0.9)  &  1.8(0.3)  & 0.99(0.01)  & 7(2)  & 3.3(0.8)  & 7.0(0.9) \\
\hline
fast/fast& 0.2 s, 0.2 s &  o3& 100& 0.79(0.16)   & 1.0(0.2)     & 0.29(0.01)  & 14(6)     & 5.0(1.2)  & 1.8(0.3)  \\
0.3\AA/cycle & & GPT5 & 100 &  0.79(0.16)   & 0.9(0.2)    &  0.28(0.01)  & 10(2)     & 4.3(0.6)   & 1.7(0.3) \\
 & 2 s, 2 s     & o3 &  100  & 1.2(0.4)        & 1.8(0.3)   & 0.30(0)   & 7.3(1.6)  & 3.3(0.8)  & 3.0(0.5) \\
 & & GPT5     & 100 & 1.7(0.4)   & 1.95(0.15)   &  0.30(0)    & 5(2)     &  2.4(0.5)   & 3.6(0.6) \\
 \hline
soft/fast& 0.2 s, 0.2 s &  o3 & 90 & 1.4(0.5) & 1.3(0.3) & 0.93(0.02) & 15(5) & 4.9(1.1) & 2.6(0.8) \\
0.3\AA/cycle & & GPT5  & 100 & 1.6(0.5) & 1.2(0.3) & 0.94(0.03) & 10(2) & 3.7(1.1) &  2.8(0.6) \\
 & 2 s, 2 s     & o3    & 100 & 2.8(0.8) & 1.8(0.3) & 0.98(0.01) & 7(2) & 3.3(0.9) &  4.7(1.0) \\
 & & GPT5       & 100 & 3.3(0.6) & 1.9(0.2) & 0.99(0.01) & 9(2) & 3.2(0.9) &  5.2(0.8) \\
 
\end{tabular}
\end{ruledtabular}
\end{table*}

\begin{figure}
\includegraphics[width=7.5cm]{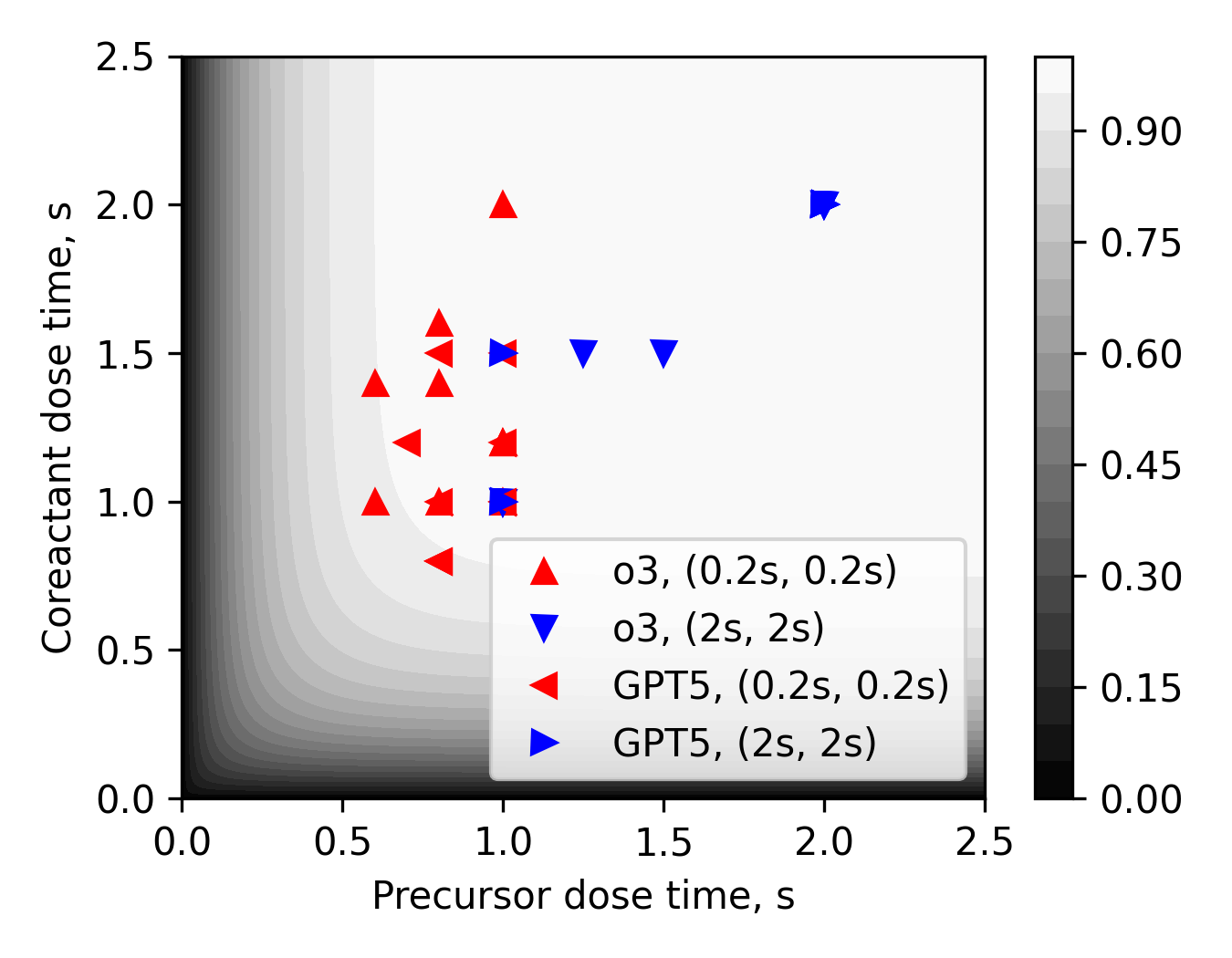}
\caption{\label{fig:scatter}  Optimized precursor and coreactant dose times for AI agents based on o3 and GPT5 models for the fast/fast process with a GPC of 
1\AA/cycle. In all cases the agents converged to an optimal solution, albeit with run-to-run variability as evidenced by the scatter in the plot. The greyscale contour
plot represents the growth per cycle as a function of the precursor and coreactant dose time}
\end{figure}

We can extract several general observations: first, except for one run involving the soft-saturating ALD process (soft/fast),
the agent reported successfully completing the process optimization in all the runs for all the ALD processes.
However, we observed significant run-to-run variability in the optimal conditions for the same
ALD process and initial guess. For instance, for the slow/slow ALD process, the standard deviation of the precursor and coreactant doses times optimized by an agent
based on the o3 reasoning model from a (0.2~s, 0.2~s) initial guess had a standard deviation of 1~s, which is more than 30\% of the average optimal dose time. 
 The initial guess also had a big impact on the optimal dose times, with the optimization starting from a (0.2~s, 0.2~s) initial guess
often resulting in significantly shorter dose times.  This is shown in Figure \ref{fig:scatter}, where we represent the optimal dose times returned by the AI agent
for the fast/fast 1~\AA/cycle ALD process. The blue dots represent runs with a (2~s,2~s) initial guess, while the red dots are those initiated with a (0.2~s, 0.2~s) guess.
There is clearly a segregation between the two conditions for both the o3 and GPT5 models, with at least two runs started using the (0.2~s, 0.2~s) guess
returning optimal dose times that lead to growth per cycles below 95\% of the saturation value.

When the initial guesses were far away
from the saturated conditions, the agent provided optimized conditions whose growth per cycle was lower than the saturated growth per cycle. For instance,
in the case of a slow/slow ALD process, optimizations started from an initial guess of (0.2~s, 0.2~s) resulted on an average GPC of 0.92 \AA/cycle and
0.86 \AA/cycle for the agents based on the o3 and GPT5 models. Similarly, optimizations from the same starting conditions of the soft/fast ALD process 
resulted in averages of 0.93 and 0.94~\AA/cycle. In contrast, for both fast/fast processes with 1.0~\AA/cycle and the slow/fast ALD process, the
average GPC of the optimal conditions equaled or exceeded 95\% that of the saturating GPC. 

\begin{figure}
\includegraphics[width=8cm]{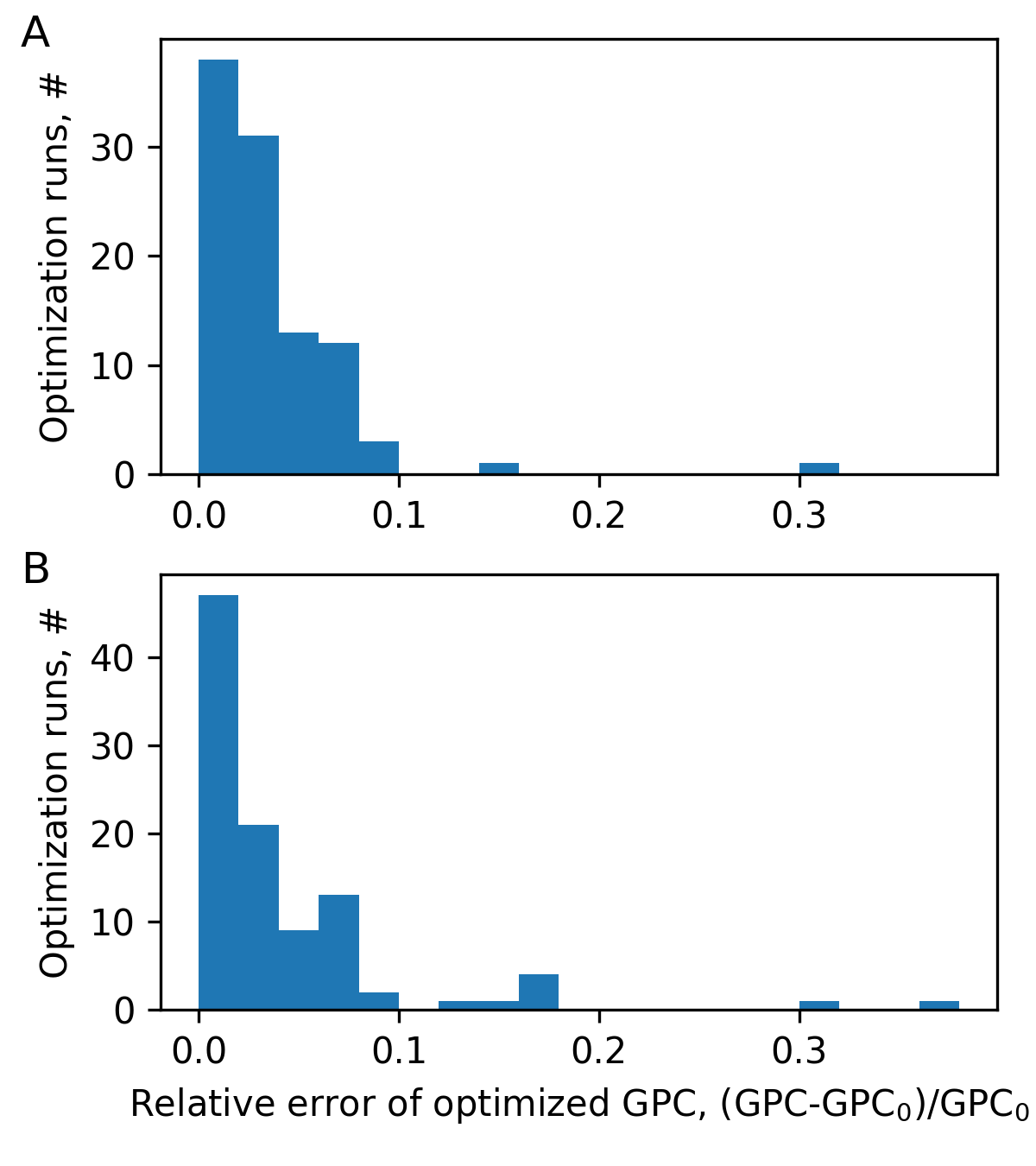}
\caption{\label{fig:histogram}Relative error between the growth per cycle resulting from the optimization process by the AI agent and the saturated GPC of the ALD process : A) 
Agent built on top of o3 reasoning model; B) Agent built on top of GPT5 reasoning model}
\end{figure}

\begin{figure}
\includegraphics[width=8cm]{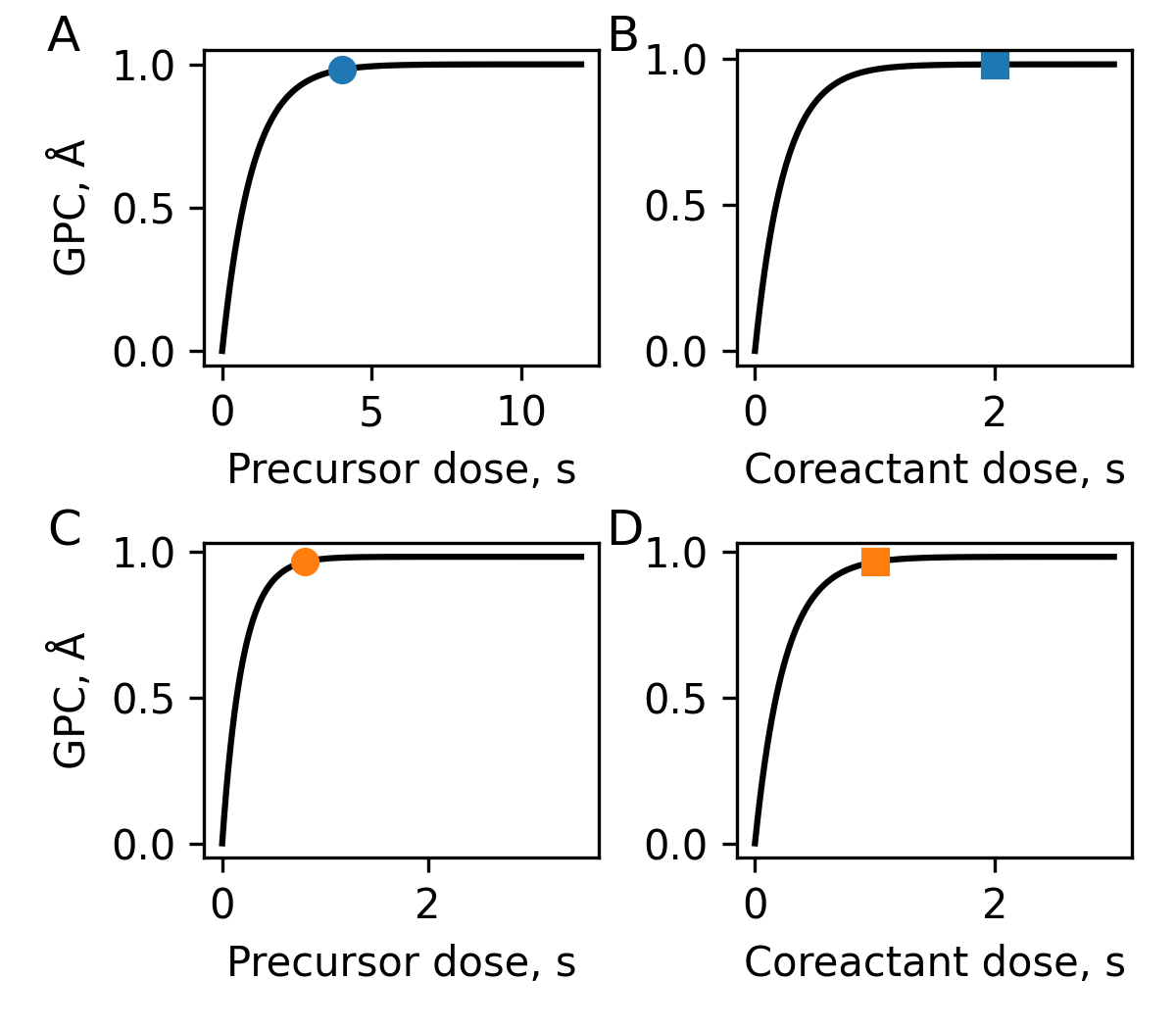}
\caption{\label{fig:fitviz}Saturation curves and optimal times for two processes optimized by an agent built with the o3 reasoning model at the the 50\% and 75\% percentile of relative error: A) precursor saturation curve and B) coreactant saturation curve for the slow/fast 1~\AA/cycle ALD process (50th percentile) C) precursor saturation curve and D) coreactant saturation curve for the fast/fast 1~\AA/cycle ALD process (75th percentile). Dots represent the optimal conditions returned by the AI agent. }
\end{figure}

If we define the relative difference between the returned GPC and
the saturated GPC, $\mathrm{GPC}_0$ as:
\begin{equation}
\varepsilon = \frac{\mathrm{GPC}_0-\mathrm{GPC}}{\mathrm{GPC}_0}
\end{equation}
we can plot the distribution of this relative difference across all the five ALD processes and initial conditions explored in this section. This results in a total of 100 points for the agent based on the o3 and GPT5 models. These distributions are shown in Figure \ref{fig:histogram}. The median relative error was 0.02 and the 75\% percentile corresponded to a relative error
of 0.05 for o3 and 0.04 for GPT5.
To help visualize the quality of the selections, we plotted selected dose times returned by the agent on top of their corresponding saturation curves. Figure \ref{fig:fitviz}
shows one example corresponding to the 50th and 75th percentiles of processes optimized by an
agent based on the o3 model. While the choice of a good set of conditions is somewhat subjective, it is clear that the AI agent tends to choose conditions that are somewhat undersaturated, particularly for R\&D work where throughput is not as critical as in a production tool.

In terms of the number of experiments required to optimize an ALD process, for most of the conditions explored the average number falls between 10 and 15. Two exceptions were the slow/slow and slow/fast ALD processes starting from a (0.2~s, 0.2~s) guess, where the average number of samples for o3 exceeded 20. A significant number of the runs starting from (2~s, 2~s) were able to find optimal dose times with an average of fewer than 10 samples. This number is almost an
order of magnitude lower than the number of samples required to optimize an ALD process using a gaussian process optimization approach described in one of our prior works\cite{Paulson2021}. However,
in that work the machine learning algorithms explored optimized both the dose and purge times of an ALD process. The number of experiments required by the AI agents is also comparable to the number of experimental points in works
reporting experimental saturation curves for both the precursor and coreactant (e. g. Refs.~\citenum{Aaltonen2005,Comstock2012,Devika2020,Hamalainen2012,Klaus2000,Klepper2007,Tero2009}). This indicates that, in terms of sample efficiency, the
performance of agents based on reasoning language models is within range of that of human experts. However, as in the case of the optimal dose times, we also observed significant run-to-run variability, as evidenced by the large standard deviation in some of the entries in Table \ref{tab:megatable}. 

\begin{figure}
\includegraphics[width=8cm]{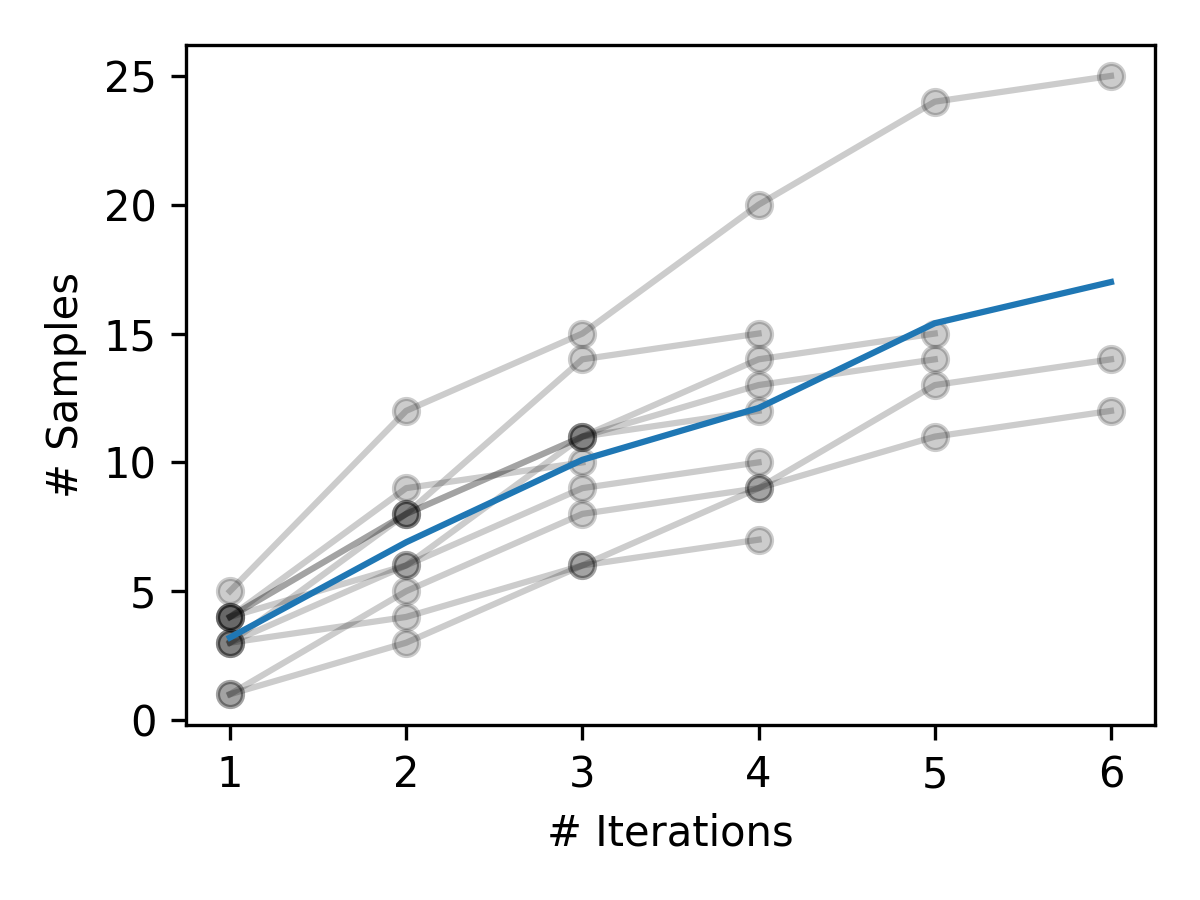}
\caption{\label{fig:samples}Number of experiments requested as a function of the iteration number for ten independent optimization runs of the fast/fast 1\AA/cycle ALD process using
an AI agent based on the o3 reasoning model for an (0.2~s, 0.2~s) initial condition. The blue line represents the average over all the runs.}
\end{figure}

From an algorithm's perspective, the number of iterations required to achieve the optimization of the fast/fast ALD process ranged between 3 and 6 or both models and starting guesses. In each
iteration, the agent usually requested between 2 and 4 new experimental conditions. In Figure \ref{fig:samples} we show how the number of samples increases during an optimization process with the
number of iterations. Each trace in Figure \ref{fig:samples} represents an independent run for the fast/fast, 1 \AA/cycle model for the (0.2~s, 0.2~s) initial guess. There is a large run-to-run variability in the way the agent suggests new samples and the total number of iterations required. In Section \ref{sec:reasoning} we will explore in more detail the 
underlying strategy reported by the agent during each iteration.

\subsection{\label{sec:noguess}Performance of base agent without initial guess}

The results obtained in Section \ref{sec:performance} show that the initial guess greatly impacts the performance of the optimization process, both in terms of the agent's ability to identify saturated conditions and the number of samples required to carry out the optimization. This motivated us
to explore how the model carries out the optimization without any initial guess and how other types of prior knowledge affected the optimization
process when supplied to the model. We focused on the slow/fast ALD process with 1.0~\AA/cycle.

We considered the following five scenarios: no prior information, a hint on the precursor pressure ("We expect the precursor to have a relatively low vapor pressure"),
a hint on the growth per cycle ("We expect the saturated growth per cycle to be close to 1.0 Angstrom/cycle"), a hint on the initial dose time ("Start exploring dose
times around 1~s"), and a scenario where we supplied both the pressure and the GPC hints. Since the differences between o3 and GPT5 are not substantial,
we carried out this analysis on the agent based on o3, which is the pure reasoning model.

\begin{table}
\caption{\label{tab:scenarios}Performance of the AI agent during the optimization of the slow/fast ALD process with no initial guess. In all cases, averages over
ten independent runs are provided, with the number in parenthesis representing the standard deviation.}
\begin{ruledtabular}
\begin{tabular}{c|ccccc}
  & Success & $t_1$ & $t_2$ & GPC & \# Samples \\
Scenario  &  (\%) & (s) & (s)  & (\AA/cy) &  \\
\hline
Baseline & 100 & 3.9(1.0) & 1.3(0.3) & 0.96(0.03) & 21(6) \\
Pressure & 100 & 3.9(0.9) & 1.5(0.4) & 0.96(0.03) & 13(3) \\
GPC & 100 & 4.0(0.7) & 1.4(0.4) & 0.97(0.02) & 16(4) \\
Both & 90 & 3.8(0.6) & 1.4(0.3)& 0.97(0.01) & 14(4) \\
Dose & 100 & 3.8(0.7) & 1.3(0.3) & 0.96 (0.02) & 14(4) \\ 
\end{tabular}
\end{ruledtabular}
\end{table}

We summarize the average dose times, number of samples, and final growth per cycle in Table \ref{tab:scenarios}. Except for one instance in the scenario where
both the pressure and GPC hints were given to the agent ("Both"), the agent reported success at
optimizing the ALD process in all the runs. The run deemed unsuccessful by the agent ocurred because in the second step the agent incorrectly interpreted the open analysis
of the reasoning model as a determination that the process was not self-limited, which terminated the optimization
process. The most salient feature in Table \ref{tab:scenarios} is the large difference in the number of samples required to optimize the ALD process for the
baseline scenario: in the absence
of a guess, the agent required an average of 21 samples to optimize the ALD process. In contrast, when additional hints were supplied to the agent, the number
decreases to the order of 14, which is a 33\% decrease. This is consistent with the underlying reasoning model in the agent leveraging this additional information
for its optimization strategy.

\begin{figure}
\includegraphics[width=7.5cm]{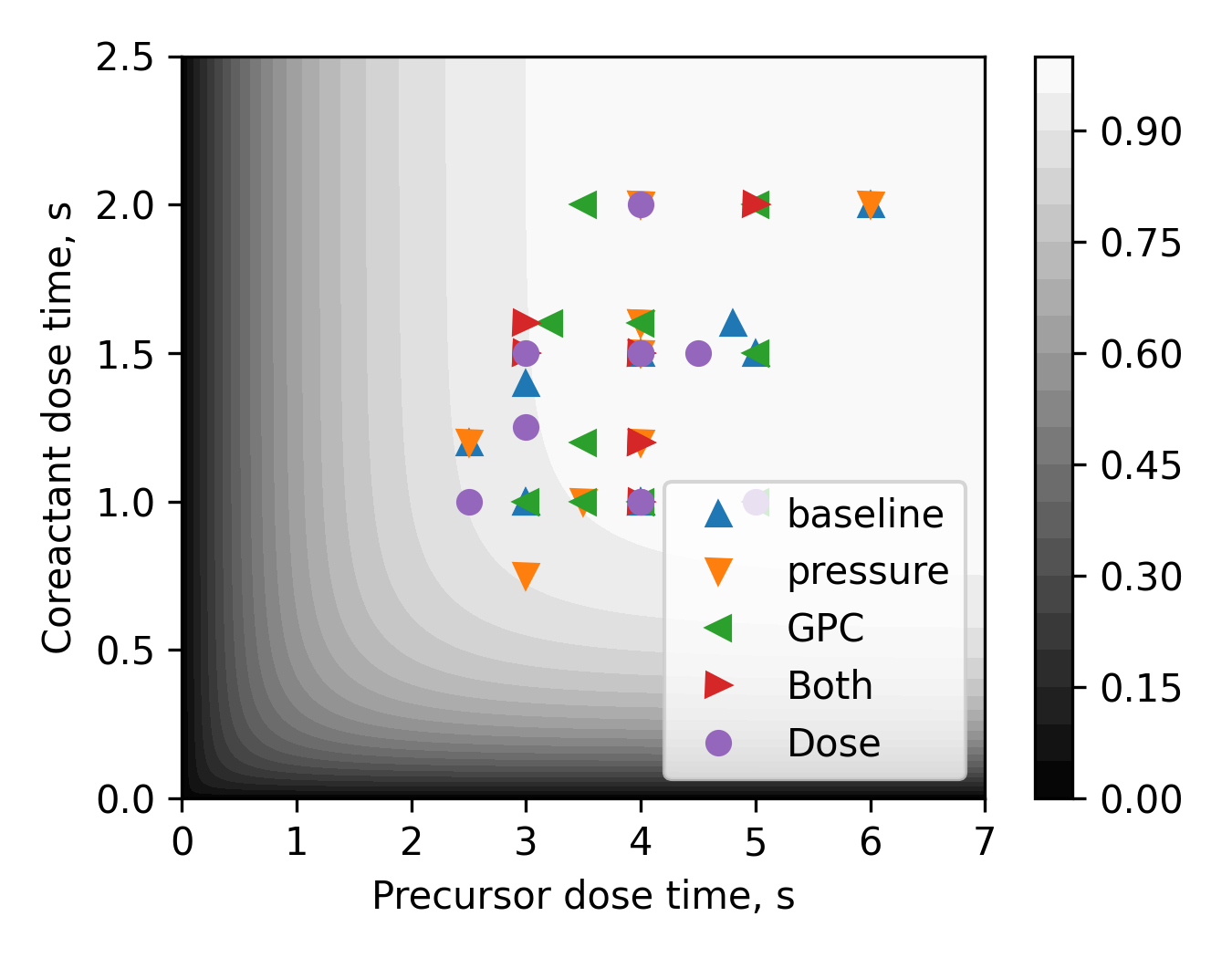}
\caption{\label{fig:scenario}  Optimized precursor and coreactant dose times for an AI agent based on the o3 model optimizing the slow/fast ALD process without
any initial guess. The different colors represent the different scenarios summarized in Table \ref{tab:scenarios}. The greyscale contour
plot represents the growth per cycle as a function of the precursor and coreactant dose time.}
\end{figure}

In Figure \ref{fig:scenario} we show a scatter plot with the optimal conditions obtained by the AI agent under the different scenarios. As in the results in Section \ref{sec:performance},
we observe a significant spread on the optimal dose times, including
a few runs where the optimal growth per cycle was 0.9 and 0.95~\AA, which
is significantly lower than the expected saturation value of 1~\AA/cycle. We do not observe any trends in terms of the optimal dose times returned by the agent for any of the scenarios considered, with the average precursor and coreactant dose times and GPC being of the same order of those in Table \ref{tab:scenarios}. This indicates that providing information to the agent primarily led
to an increase in sample efficiency and not a significant change in the outcome of the optimization process.

Finally, we evaluated the agent's ability to identify the presence of a CVD component as it attempts the process optimization. We considered the fast/fast 1\AA/cycle ALD process and we added a CVD component as described in Section \ref{sec:aldmodel}. This
results in a linear dependence of the growth per cycle with precursor dose time (see Fig.~\ref{fig:sat}B). We asked the agent to
optimize this process without any additional information. Then, we calculated the percentage of runs that ended with the agent concluding that the process is not fully self-limited.

We carried out this study for increasing magnitudes of the CVD component. For a CVD component of 3~\AA/min, only 3 out of 10 runs concluded that the process was not self-limited. As the magnitude of the CVD component increases, the probability that the agent concludes that the process is not self-limited also increases, with 6 out of 10 runs and 5 out of 10 runs correctly identifying that the process was not self-limited for CVD components of 4.8~\AA/min and 6~\AA/min, respectively. The run-to-run variability observed in the fully self-limited ALD processes therefore extends to the determination of whether a new process is actually self-limited.

\subsection{\label{sec:memory}Performance of agents with memory}

\begin{table}
\caption{\label{tab:memory}Performance of the AI agent with memory during the optimization of the slow/fast ALD process with no initial guess. The agent uses the o3 reasoning language model. In all cases, averages over ten independent runs are provided, with the number in parenthesis representing the standard deviation.}
\begin{ruledtabular}
\begin{tabular}{c|ccccc}
  & Success & $t_1$ & $t_2$ & GPC & \# Samples \\
Scenario  &  (\%) & (s) & (s)  & (\AA/cy) &  \\
\hline
Baseline & 100 & 3.7(0.8) & 1.0(0.2) &  0.95(0.02) & 18(5) \\
Pressure & 100 &  3.9(1.1) & 1.0(0.2) & 0.94(0.04) & 16(4) \\
GPC & 100 &  3.7(0.7) & 1.1(0.4) & 0.94(0.06) & 18(4) \\
Both & 100 & 4.2(1.0) & 1.1(0.2) & 0.96(0.02) & 18(2) \\
Dose & 100 & 3.9(0.5) & 1.1(0.3) & 0.96(0.02) & 13.6(1.4) \\ 
\end{tabular}
\end{ruledtabular}
\end{table}

\begin{figure}
\includegraphics[width=7.5cm]{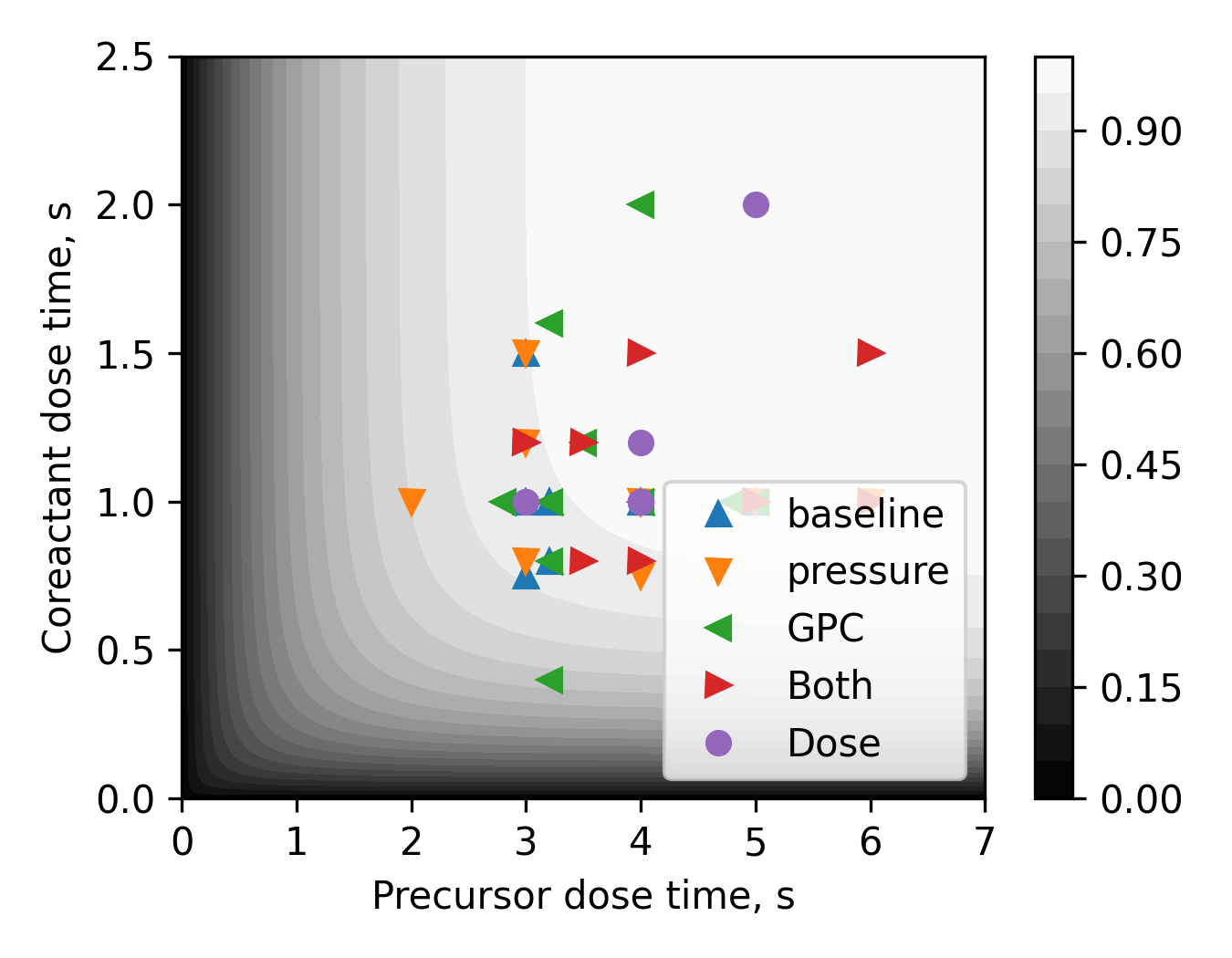}
\caption{\label{fig:memory}  Optimized precursor and coreactant dose times for an AI agent with memory based on the o3 model optimizing the slow/fast ALD process without
any initial guess. The different colors represent the different scenarios summarized in Table \ref{tab:memory}. The greyscale contour
plot represents the growth per cycle as a function of the precursor and coreactant dose time.}
\end{figure}

To understand if the performance of the agents improved by providing information on the model's reasoning in past iterations, we carried out optimization runs based on the o3 reasoning models using the same conditions explored in Section \ref{sec:noguess}.  The results, summarized in Table \ref{tab:memory}, did not show any major differences. We did observe an overall decrease in the number of experiments required when no hints were supplied and when an estimate of the dose time was supplied to the agent
(baseline and dose scenarios in both Table \ref{tab:scenarios} and \ref{tab:memory}). 

In contrast, the average growth per cycle obtained at the end of the optimization process was
slightly lower, which resulted in some runs providing optimal dose times that were not saturated (Figure \ref{fig:memory}).

This shows that passing the additional information to the
underlying model to maintain consistency in the optimization process does not lead to significant improvements in the agent's performance.

\subsection{\label{sec:reasoning}Model search strategy and reasoning}

For all the scenarios described above, we have compiled both the sequence of experiments requested per iteration and the long form responses from the reasoning language model supporting each request.  We can use this data to explore in more detail the optimization strategy of agents powered by reasoning models. A
simple way of visualizing this strategy is representing each experiment request as a scatter plot, where each point is colored according to the iteration number
where it was requested.

\begin{figure*}
\includegraphics[width=16cm]{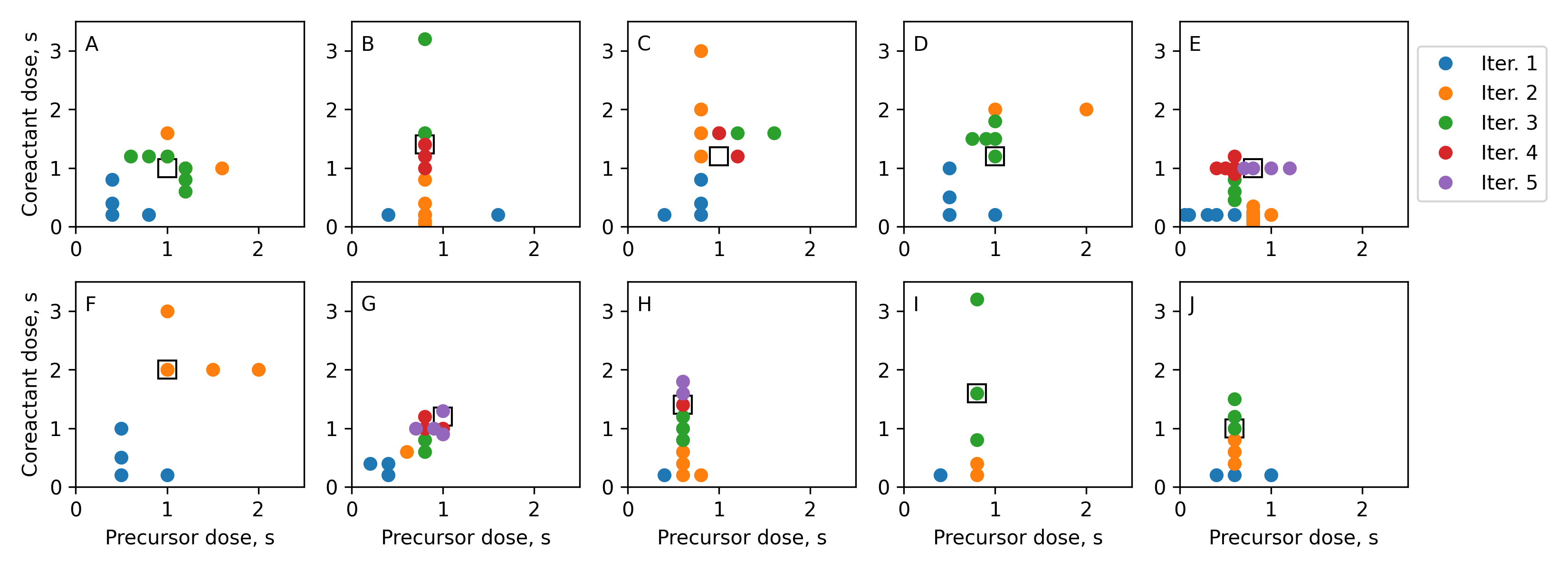}
\caption{\label{fig:strategy1}Visualization of the search
strategy of the baseline agent build on top of OpenAI's o3
reasoning model
when optimizing a fast/fast ALD process with a GPC of
1~\AA/cycle with an initial guess of (0.2~s, 0.2~s). Each color represents a request for additional
data points made by the agent during a different iteration of
the optimization process. Plots A-J each represent an independent run, showcasing the variability in the search strategy of the agent.}
\end{figure*}

\begin{figure*}
\includegraphics[width=16cm]{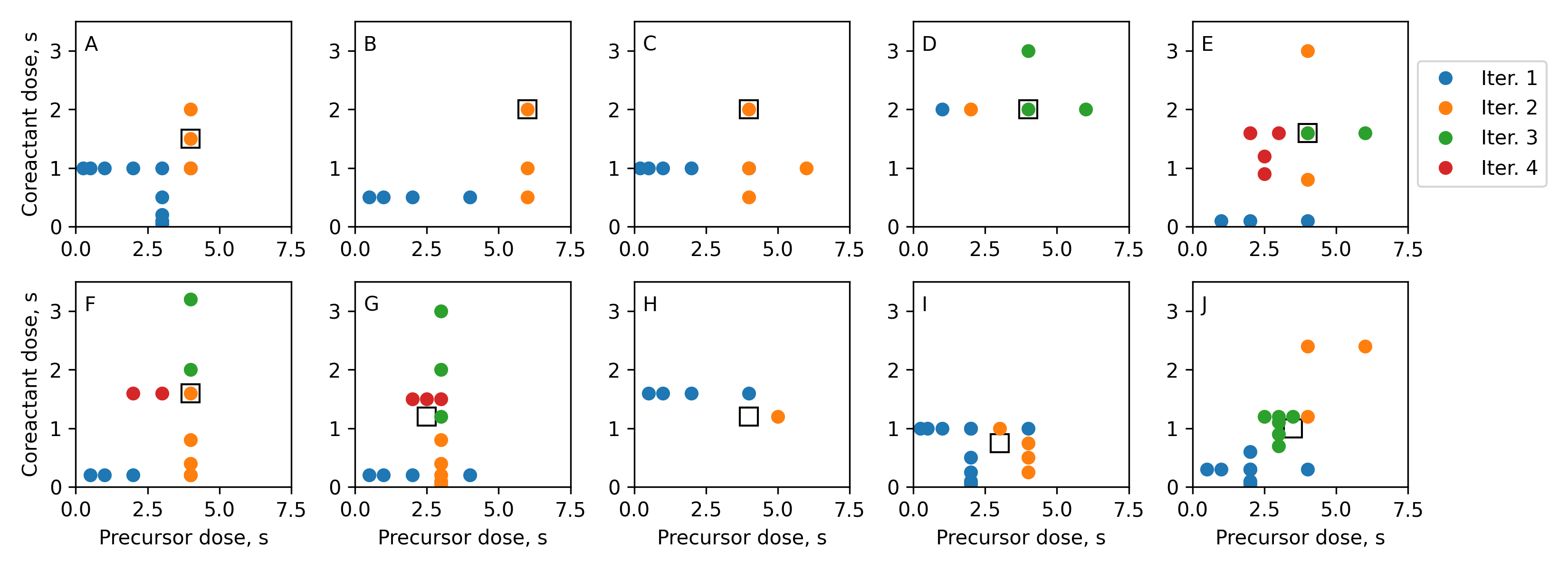}
\caption{\label{fig:strategy10}Visualization of the search
strategy of the baseline agent built on top of OpenAI's o3 
reasoning model
when optimizing a slow/fast ALD process with a GPC of
1~\AA/cycle with no initial guess and with prior information
on the precursor vapor pressure. Plots A-J each represent an independent run.}
\end{figure*}

We have analyzed these plots to identify a few salient strategies. We can then use the long-form responses to gain insights into the chain of thought that the reasoning
model uses for each of the strategies. In Figure \ref{fig:strategy1} we present a representation of the experimental requests as an agent built on top of the o3 model tries to optimize the fast/fast 1.0 \AA/cycle process with
an initial guess of (0.2~s,0.2~s). In this representation, the black dot represents the initial guess, and subsequent requests are colored by iteration. The final optimized condition is indicated by a square. 
Each of the plots in Fig \ref{fig:strategy1} represents one independent
optimization attempt by the model. Note that some of the squares in Fig.~\ref{fig:strategy1} do not overlap with prior data points. This indicates that the optimal condition
has not been experimentally validated and it is a new proposal from the model.

The first salient feature is the lack of a unique strategy: in
some instances, such as Fig.~\ref{fig:strategy1}(A), (D), (F), and (G) the agent starts exploring regions near the initial guess and then moves to higher dose times.
In other instances, such as Fig. \ref{fig:strategy1}(B), (H), (I) and (J) the optimization clearly moves along 1D saturation curves. Finally, the optimization runs
captured in Fig. Fig. \ref{fig:strategy1}(C) and (E) present a hybrid strategy, where exploration takes place primarily along saturation curves, some of which
were interrupted before moving into a different part of the configuration space. There is also significant variability in terms of whether the final proposal from the agent has been experimentally validated, as evidenced by the square not overlapping existing data points. For this specific case, 3 out of the 10 runs
captured in Fig. \ref{fig:strategy1} yielded optimal dose times that had not been experimentally validated.

The agents develop their strategy across multiple iterations: the number of exploration points requested during the first iteration was small, approximately three. We observed a similar behavior across models and ALD processes when a guess is provided. This may be one source of variability in the agent's strategy:
in the agent configuration used to obtain the results summarized in Fig. \ref{fig:strategy1}, the reasoning model is shown just the available data during each iteration, and not the thought process during the previous iterations. This may lead to changes in strategy simply because the model is reacting to the experimental data and is not trying to follow a plan set during the first iteration.

In Figure \ref{fig:strategy10} we show a similar representation for the optimization of the slow/fast ALD process with 1~\AA/cycle in the absence of an initial guess and with additional information on the precursor vapor pressure. The overall number of experiments required is similar than in Fig. \ref{fig:strategy1}. We can see the same overall strategies in this case. Also, 4 out of 10 optimal conditions have not been experimentally verified. However, a significant difference compared to optimizations starting from an externally supplied guess is the variability in the initial experiments requested by the agents (blue dots in Fig. \ref{fig:strategy10}). In terms of the number of requests, the majority of examples in Fig. \ref{fig:strategy10} include 4 or more requests in the first iteration. Whenever the number of points was greater than one, the reasoning model favored primarily exploring the saturation behavior of the precursor (e.g. Fig. \ref{fig:strategy10}(B), (C), and (E) to (H)). In some of the runs, such as those represented in Fig. \ref{fig:strategy10}(A), (I) and (J), the reasoning model returns requests for saturation curves for both the precursor and the co-reactant. The dose time for the coreactant during this first exploration showed a marked run-to-run variability as well.

When we look at the open response of the reasoning model, we observe
that the model's responses consistently show that they are grounded
in the physics of self-limited. process. For instance, the following is
an excerpt of the agent's conclusion after the first iteration of an
optimization process:

\begin{quote}
Outcome: If both precursor and coreactant show a clear plateau, the process is self-limited and optimized at those shortest plateau doses (likely $\sim$0.40~s precursor, $\sim$0.40~s coreactant).  If GPC keeps increasing with dose without a plateau, the reaction is not self-limited and a different chemistry or temperature window must be considered.
\end{quote}

Likewise, during the last iteration, the agent justifies ending the optimization
process as follows:

\begin{quote}
With a generous coreactant dose ($\ge 1$~s) the GPC rises from 0.94~\AA (0.6~s precursor) 
$\rightarrow$ 0.97~\AA (0.8~s) $\rightarrow$ 0.99~\AA(1.0~s) and then stays constant at 0.98-0.99~\AA for 1.2 s and 1.6~s precursor. Therefore the precursor dose is saturated at about 1.0~s; anything longer offers no measurable gain.
\end{quote}

The responses consistently showed this type of reasoning. 

Finally,
we can also look at the agent's reasoning when the process
has an additional CVD component. This is a snapshot of the reasoning leading the model to conclude that the fast/fast
1\AA/cycle process with a CVD component of 5.0 \AA/cycle is not self-limited:

\begin{quote}

When the coreactant is already saturated, the growth continues to climb monotonically from 1.03 \AA/cycle (precursor 0.8 s) to 2.46 \AA/cycle (precursor 15 s).  No hint of a plateau is visible, even at 15 s. The rise is roughly proportional to dose, which is contrary to the self-limiting behaviour expected for ALD. Conclusion: The process is NOT optimized.  The precursor half-cycle is clearly not saturated and, given the very long doses already explored, the chemistry is probably not self-limited under the present conditions.

\end{quote}

\begin{figure}
\includegraphics[width=7.5cm]{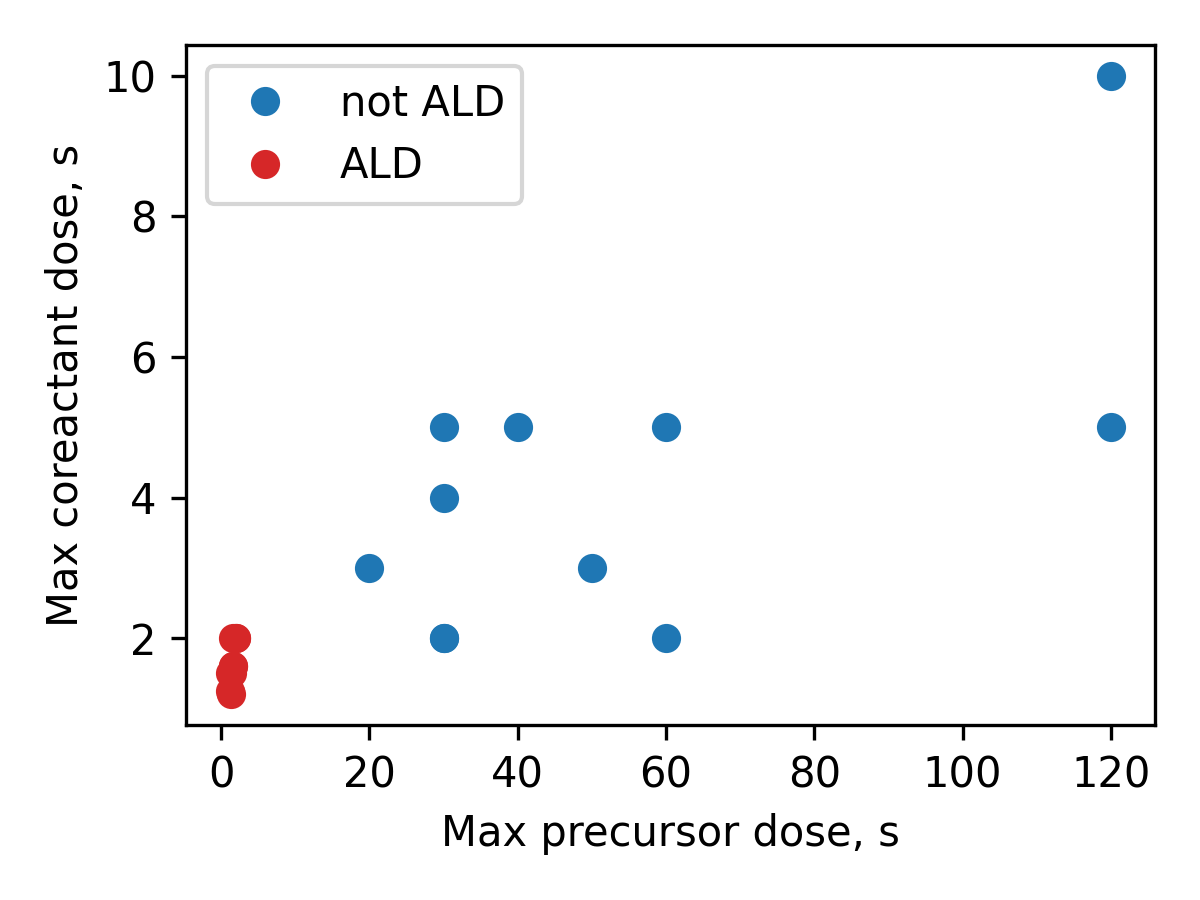}
\caption{\label{fig:exploration}  Maximum precursor and
dose times explored by the AI agent based on the o3 model for
two fast/fast, 1~\AA/cycle ALD processes with CVD components
of 4.8~\AA/min and 6.0~\AA/min. Each point represents a 
separate run and they are color coded depending on whether
the agent identifies the process as self-limited (red) or
non self-limited (blue)}
\end{figure}

This shows that the model is capable of using a reasoning that is largely correct. To gain further insight into how the agent's choices during the optimization can affect or bias their reasoning we focused on the ALD processes with a non self-limited CVD component studied in Section \ref{sec:noguess}. Figure \ref{fig:exploration} shows the largest dose times explored by the agent for independent runs of the agent built on o3 reasoning models for
the fast/fast ALD process for two of the values for
the CVD component explored in Section \ref{sec:noguess}:
4.8~\AA/min and 6.0~\AA/min. Each point is a separate run and
they are colored depending on whether the agent determined
that the process was self-limited (blue) or non self-limited (red).

Figure \ref{fig:exploration} shows a clear correlation
between the range of dose times explored and whether the
agent was able to correctly identify the process as a non self-limited process. This confirms that the variability in the results does not seem to stem from reasoning issues, but from how the reasoning process is biased by the agent's own choices when exploring the design space.

\section{Discussion}

Agents based on reasoning models are capable of optimizing ALD processes without any prior knowledge of the growth per cycle and saturation times. In the majority of cases,
they were capable of converging to reasonable dose times using a number of experiments close to those that would be used by human experts optimizing a new ALD process. A
deeper analysis, though, showed a strong run-to-run variability in the agent's performance. Moreover, agents consistently
struggled to identify non self-limiting conditions without
any additional priors such as the expected saturation
GPC. This was true for both the o3 and GPT5 models.

Both models used in this work are reasoning
language models, more advanced models built on top of conventional LLMs that are capable of internally breaking down complex
problems and use multiple calls to LLMs to work on individual steps and integrate the final results. When we analyzed the underlying reasoning process used by o3 and GPT5, 
the reasoning traces showed that the agents were using relevant concepts such as saturation plateaus, or making decisions based on the dependence of growth per cycle with dose times, that are well aligned with how human experts reason about ALD processes. However, we also saw a high degree of variability in the strategies used by the agents regardless of whether the agent had access to past reasoning steps or started with just a prior set of growth conditions at each iteration. For the case of the non self-limited
processes, we have seen that it was the search over the parameter
space that biased the agent's determination of the self-limited
nature of an ALD process. 

However, we should note that we have purposely focused on
the worst case scenario for these agents: not only do the
agents not have any prior information about the ALD process,
but the prompt to the model in charge to the optimization
process lacks details (see Appendix \ref{sec:prompt}). The
performance reported here is therefore that of a baseline agent.
As we have seen, the mere addition
of hints reduces the number of samples required to optimize
the process. Consequently, refinements in both the prompt
and agent design will likely lead to 
significant improvements.
In order to facilitate this exploration,
the ALD model is available as open source in a repository.

Based on these observations, how useful are agents based on reasoning models in practice? One of the most promising use cases for these agents is the optimization of new processes with real time feedback from in-situ characterization techniques. The results show that in most conditions agents can explore the parameter space in a fully unsupervised way.  If we used them exclusively for fully automated optimization tasks on self-limited processes, their advantages with respect to either rule-based algorithms or machine learning approaches are not significant. However, one
 unique capability is their ability to respond to different
types of text-based requests.

From an experimental standpoint, the integration of these agents with ALD tools is straightforward. In a separate work, we describe how we have augmented an existing ALD tool with AI capabilities, including agents based on LLMs. We were able to show that the AI component does not slow down the IO operations required to monitor and control an ALD reactor. One
important consideration in experimental systems, particularly when real time feedback is used, is the lag between when a query or command is issued to the agent and the actual start of a new ALD growth. Experimentally, we have seen that this is dominated by the time the agent takes to process the request rather than the communication with the ALD reactor. In our specific case, this lag
was always a few seconds per iteration, which makes agents based
on reasoning models
suitable for real time process optimizations using in-situ techniques. Details on the performance of the experimental
system will be reported elsewhere.

\section{Conclusions}

In this work, we have demonstrated that AI agents based on reasoning large language models such as OpenAI's o3 and GPT5 can successfully optimize ALD processes in a fully autonomous manner without prior knowledge of the process parameters. These agents consistently identified optimal dose times using a number of experiments comparable to those employed by human experts, typically requiring 10-15 samples for most processes. Analysis of the reasoning traces revealed that the models employ sound logic grounded in fundamental ALD concepts such as saturation plateaus and self-limiting behavior.  However, we observed significant run-to-run variability in both the optimization strategies and the final dose times selected, with the agents' own choices during parameter space exploration influencing their conclusions. The agents also struggled to reliably identify non-self-limiting processes. While the baseline agent architecture presented here operates in a worst-case scenario with minimal prompting and no process-specific priors, the addition of even simple hints substantially improved sample efficiency, suggesting that refined prompt engineering and agent design could yield significant performance gains.

\begin{acknowledgments}
This research is based upon work supported by Laboratory Directed Research and Development (LDRD) funding from Argonne National Laboratory, provided by the Director, Office of Science, of the U.S. Department of Energy under Contract No. DE-AC02-06CH11357.

\end{acknowledgments}

\section*{Data Availability Statement}

Data available in article or supplementary material.

\appendix

\section{\label{sec:appendixmodel}Derivation of ALD model}

If we define $\theta$ as the fraction of surface sites
that have reacted with an ALD precursor, a simple irreversible
Langmuir kinetics model established the following equation
for the evolution of the surface coverage as a function of time:
\begin{equation}
\frac{d\theta}{dt} = k_1(1-\theta)
\end{equation}
Where the constant $k_1$ incorporates the rate coefficient as 
well as the dependence with the precursor pressure. We can
add a non self-limited component by considering an effective
concentration of coreactant during the precursor dose, $k_c$,
so that the kinetic model is now:
\begin{equation}
\label{eq:aldcvd}
\frac{d\theta}{dt} = k_1(1-\theta) - k_c\theta
\end{equation}
The growth per cycle is therefore given by:
\begin{equation}
\label{eq:growth}
\mathrm{GPC} = \mathrm{GPC_0} \int_0^{t_1} k_1(1-\theta) dt
\end{equation}

Likewise, if $\theta_1$ is the fractional surface coverage
at the end of the dose time, assuming ideal purge times the
evolution of the surface coverage during the coreactant
dose is:
\begin{equation}
\frac{d\theta}{dt} = - k_2\theta
\end{equation}
Which has the simple solution:
\begin{equation}
\label{eq:aldcvd2}
\theta_2 = \theta_1 e^{-k_2t_2}
\end{equation}
Here $t_2$ is the duration of the coreactant dose and we define
$\theta_2$ as the fractional coverage at the end of the coreactant
dose.

Eq. \ref{eq:aldcvd} can be trivially solved to obtain $\theta_1$
as a function of $\theta_0$:
\begin{equation}
\label{eq:precpulse}
\theta(t) = \theta_\mathrm{lim}\left(1-e^{-(k_1+k_c)t}\right) + \theta_0 e^{-(k_1+k_c)t}
\end{equation}
where
\begin{equation}
\theta_\mathrm{lim} = \frac{k_1}{k_1+k_c}
\end{equation}
From Eq. \ref{eq:precpulse} we can relate the asymptotic
non self-limited growth rate $\mathrm{GR}_0$ and the kinetic
constant $k_c$:
\begin{equation}
\mathrm{GR}_0 = \mathrm{GPC}_0\frac{k_c}{k_1+k_c}
\end{equation}
We can also obtain $\theta_1$ as $\theta_1 = \theta(t_1)$ in Eq. \ref{eq:precpulse}.

In an steady state ALD process, we have that:
\begin{equation}
\theta_0 = \theta_2
\end{equation}
This provides an equation to solve for $\theta_0$ as a function
of the precursor and coreactant dose times $t_1$ and $t_2$. Using
Eq. \ref{eq:growth} we obtain the expressions used in
Section \ref{sec:aldmodel}, including the particular
case of a fully self-limited model for which $k_c=0$.

This approach can be trivially extended to multiple reaction pathways for both the precursor and the coreactant.

\section{\label{sec:prompt}Prompts used for the reasoning model}

Here we provide verbatim the prompt used for the reasoning
model during each iteration:

\begin{verbatim}
## Your job: process requests to operate an
atomic layer deposition process.

You are in charge of optimizing an atomic
layer deposition process. 

Atomic layer deposition (ALD) is a thin film
technique where a given process is
characterized by four times: the dose time
for the precursor, the purge time for the
precursor, the dose time for the coreactant,
and the purge time for the coreactant.

ALD is self-limited: for long enough dose
times the growth per cycle becomes saturated.

Your job is to determine if the process is
already optimized based on the data provided
and, if it is not saturated, provide some new
experimental conditions to try.

Also, at some point if the dose times are
too long and the growth rate keeps increasing,
you may conclude that the process is not
self-limited.

You only have to provide the dose times for
the precursor and the coreactant. The purge
times have already been optimized. 

Remember that too long of a dose time is
wasteful both in terms of precursor
utilization and the process duration.
Therefore, you have to find dose times for
the precursor and co-reactant that are
large enough to be saturated but not too
long so that there is significant waste.

Remember that all self-limited process start
with a strong dependence of the growth per
cycle with dose time until the growth becomes
saturated.

Remember that a process may be saturated for
the precursor dose, but not saturated for the
co-reactant. It is therefore paramount to
check both.

The data provided will contain any useful
prior information and a list of conditions
listing the precursor dose time ("precursor"),
coreactant dose time ("coreactant"), and the
corresponding growth per cycle ("gpc")

In some cases no prior data will be available
and you will have to provide initial guesses
for the dose and purge times. In other cases
you will receive a specific request in terms
of the optimization strategy. 

\end{verbatim}

During each iteration, the prior growth conditions were
appended to this prompt to generate the query that was
passed to the model.

\bibliography{llm_paper, saturation_curves}

@article{YanguasGil2014,
	author = {Yanguas-Gil, Angel and Elam, Jeffrey W.},
	journal = {Journal of Vacuum Science \& Technology A},
	month = {03},
	number = {3},
	pages = {031504},
	title = {Analytic expressions for atomic layer deposition: Coverage, throughput, and materials utilization in cross-flow, particle coating, and spatial atomic layer deposition},
	volume = {32},
	year = {2014}}

@article{Yanguasgil2021,
	author = {Yanguas-Gil, Angel and Libera, Joseph A. and Elam, Jeffrey W.},
	journal = {Journal of Vacuum Science \& Technology A},
	month = {09},
	number = {6},
	pages = {062404},
	title = {Reactor scale simulations of {ALD} and {ALE}: Ideal and non-ideal self-limited processes in a cylindrical and a 300 mm wafer cross-flow reactor},
	volume = {39},
	year = {2021}}

@misc{zhang2025,
	archiveprefix = {arXiv},
	author = {Jie Zhang and Cezara Petrui and Kristina Nikoli{\'c} and Florian Tram{\`e}r},
	eprint = {2505.12575},
	primaryclass = {cs.AI},
	title = {RealMath: A Continuous Benchmark for Evaluating Language Models on Research-Level Mathematics},
	url = {https://arxiv.org/abs/2505.12575},
	year = {2025}}

@misc{balunovic2025,
	archiveprefix = {arXiv},
	author = {Mislav Balunovic and Jasper Dekoninck and Ivo Petrov and Nikola Jovanovic and Martin Vechev},
	eprint = {2505.23281},
	primaryclass = {cs.AI},
	title = {MathArena: Evaluating LLMs on Uncontaminated Math Competitions},
	url = {https://arxiv.org/abs/2505.23281},
	year = {2025}}

@misc{Wei2023,
	archiveprefix = {arXiv},
	author = {Jason Wei and Xuezhi Wang and Dale Schuurmans and Maarten Bosma and Brian Ichter and Fei Xia and Ed Chi and Quoc Le and Denny Zhou},
	eprint = {2201.11903},
	primaryclass = {cs.CL},
	title = {Chain-of-Thought Prompting Elicits Reasoning in Large Language Models},
	url = {https://arxiv.org/abs/2201.11903},
	year = {2023}}

@misc{Yang2025,
	archiveprefix = {arXiv},
	author = {An Yang and Anfeng Li and Baosong Yang and Beichen Zhang and Binyuan Hui and Bo Zheng and Bowen Yu and Chang Gao and Chengen Huang and Chenxu Lv and Chujie Zheng and Dayiheng Liu and Fan Zhou and Fei Huang and Feng Hu and Hao Ge and Haoran Wei and Huan Lin and Jialong Tang and Jian Yang and Jianhong Tu and Jianwei Zhang and Jianxin Yang and Jiaxi Yang and Jing Zhou and Jingren Zhou and Junyang Lin and Kai Dang and Keqin Bao and Kexin Yang and Le Yu and Lianghao Deng and Mei Li and Mingfeng Xue and Mingze Li and Pei Zhang and Peng Wang and Qin Zhu and Rui Men and Ruize Gao and Shixuan Liu and Shuang Luo and Tianhao Li and Tianyi Tang and Wenbiao Yin and Xingzhang Ren and Xinyu Wang and Xinyu Zhang and Xuancheng Ren and Yang Fan and Yang Su and Yichang Zhang and Yinger Zhang and Yu Wan and Yuqiong Liu and Zekun Wang and Zeyu Cui and Zhenru Zhang and Zhipeng Zhou and Zihan Qiu},
	eprint = {2505.09388},
	primaryclass = {cs.CL},
	title = {Qwen3 Technical Report},
	url = {https://arxiv.org/abs/2505.09388},
	year = {2025}}

@article{Guo_2025,
	author = {Guo, Daya and Yang, Dejian and Zhang, Haowei and Song, Junxiao and Wang, Peiyi and Zhu, Qihao and Xu, Runxin and Zhang, Ruoyu and Ma, Shirong and Bi, Xiao and Zhang, Xiaokang and Yu, Xingkai and Wu, Yu and Wu, Z. F. and Gou, Zhibin and Shao, Zhihong and Li, Zhuoshu and Gao, Ziyi and Liu, Aixin and Xue, Bing and Wang, Bingxuan and Wu, Bochao and Feng, Bei and Lu, Chengda and Zhao, Chenggang and Deng, Chengqi and Ruan, Chong and Dai, Damai and Chen, Deli and Ji, Dongjie and Li, Erhang and Lin, Fangyun and Dai, Fucong and Luo, Fuli and Hao, Guangbo and Chen, Guanting and Li, Guowei and Zhang, H. and Xu, Hanwei and Ding, Honghui and Gao, Huazuo and Qu, Hui and Li, Hui and Guo, Jianzhong and Li, Jiashi and Chen, Jingchang and Yuan, Jingyang and Tu, Jinhao and Qiu, Junjie and Li, Junlong and Cai, J. L. and Ni, Jiaqi and Liang, Jian and Chen, Jin and Dong, Kai and Hu, Kai and You, Kaichao and Gao, Kaige and Guan, Kang and Huang, Kexin and Yu, Kuai and Wang, Lean and Zhang, Lecong and Zhao, Liang and Wang, Litong and Zhang, Liyue and Xu, Lei and Xia, Leyi and Zhang, Mingchuan and Zhang, Minghua and Tang, Minghui and Zhou, Mingxu and Li, Meng and Wang, Miaojun and Li, Mingming and Tian, Ning and Huang, Panpan and Zhang, Peng and Wang, Qiancheng and Chen, Qinyu and Du, Qiushi and Ge, Ruiqi and Zhang, Ruisong and Pan, Ruizhe and Wang, Runji and Chen, R. J. and Jin, R. L. and Chen, Ruyi and Lu, Shanghao and Zhou, Shangyan and Chen, Shanhuang and Ye, Shengfeng and Wang, Shiyu and Yu, Shuiping and Zhou, Shunfeng and Pan, Shuting and Li, S. S. and Zhou, Shuang and Wu, Shaoqing and Yun, Tao and Pei, Tian and Sun, Tianyu and Wang, T. and Zeng, Wangding and Liu, Wen and Liang, Wenfeng and Gao, Wenjun and Yu, Wenqin and Zhang, Wentao and Xiao, W. L. and An, Wei and Liu, Xiaodong and Wang, Xiaohan and Chen, Xiaokang and Nie, Xiaotao and Cheng, Xin and Liu, Xin and Xie, Xin and Liu, Xingchao and Yang, Xinyu and Li, Xinyuan and Su, Xuecheng and Lin, Xuheng and Li, X. Q. and Jin, Xiangyue and Shen, Xiaojin and Chen, Xiaosha and Sun, Xiaowen and Wang, Xiaoxiang and Song, Xinnan and Zhou, Xinyi and Wang, Xianzu and Shan, Xinxia and Li, Y. K. and Wang, Y. Q. and Wei, Y. X. and Zhang, Yang and Xu, Yanhong and Li, Yao and Zhao, Yao and Sun, Yaofeng and Wang, Yaohui and Yu, Yi and Zhang, Yichao and Shi, Yifan and Xiong, Yiliang and He, Ying and Piao, Yishi and Wang, Yisong and Tan, Yixuan and Ma, Yiyang and Liu, Yiyuan and Guo, Yongqiang and Ou, Yuan and Wang, Yuduan and Gong, Yue and Zou, Yuheng and He, Yujia and Xiong, Yunfan and Luo, Yuxiang and You, Yuxiang and Liu, Yuxuan and Zhou, Yuyang and Zhu, Y. X. and Huang, Yanping and Li, Yaohui and Zheng, Yi and Zhu, Yuchen and Ma, Yunxian and Tang, Ying and Zha, Yukun and Yan, Yuting and Ren, Z. Z. and Ren, Zehui and Sha, Zhangli and Fu, Zhe and Xu, Zhean and Xie, Zhenda and Zhang, Zhengyan and Hao, Zhewen and Ma, Zhicheng and Yan, Zhigang and Wu, Zhiyu and Gu, Zihui and Zhu, Zijia and Liu, Zijun and Li, Zilin and Xie, Ziwei and Song, Ziyang and Pan, Zizheng and Huang, Zhen and Xu, Zhipeng and Zhang, Zhongyu and Zhang, Zhen},
	doi = {10.1038/s41586-025-09422-z},
	issn = {1476-4687},
	journal = {Nature},
	month = sep,
	number = {8081},
	pages = {633--638},
	publisher = {Springer Science and Business Media LLC},
	title = {DeepSeek-R1 incentivizes reasoning in LLMs through reinforcement learning},
	url = {http://dx.doi.org/10.1038/s41586-025-09422-z},
	volume = {645},
	year = {2025}}

@article{Klaus2000,
	author = {J.W Klaus and S.J Ferro and S.M George},
	doi = {https://doi.org/10.1016/S0040-6090(99)01076-7},
	issn = {0040-6090},
	journal = {Thin Solid Films},
	number = {1},
	pages = {145-153},
	title = {Atomic layer deposition of tungsten using sequential surface chemistry with a sacrificial stripping reaction},
	url = {https://www.sciencedirect.com/science/article/pii/S0040609099010767},
	volume = {360},
	year = {2000}}

@article{Hamalainen2012,
	author = {H{\"a}m{\"a}l{\"a}inen, Jani and Sajavaara, Timo and Puukilainen, Esa and Ritala, Mikko and Leskel{\"a}, Markku},
	date = {2012/01/10},
	doi = {10.1021/cm201795s},
	isbn = {0897-4756},
	journal = {Chemistry of Materials},
	journal1 = {Chemistry of Materials},
	journal2 = {Chem. Mater.},
	month = {01},
	number = {1},
	pages = {55--60},
	publisher = {American Chemical Society},
	title = {Atomic Layer Deposition of Osmium},
	type = {doi: 10.1021/cm201795s},
	url = {https://doi.org/10.1021/cm201795s},
	volume = {24},
	year = {2012},
	year1 = {2012}}

@article{Paulson2021,
	author = {Paulson, Noah H. and Yanguas-Gil, Angel and Abuomar, Osama Y. and Elam, Jeffrey W.},
	date = {2021/04/14},
	doi = {10.1021/acsami.1c00649},
	isbn = {1944-8244},
	journal = {ACS Applied Materials \& Interfaces},
	journal1 = {ACS Applied Materials \& Interfaces},
	journal2 = {ACS Appl. Mater. Interfaces},
	month = {04},
	number = {14},
	pages = {17022--17033},
	publisher = {American Chemical Society},
	title = {Intelligent Agents for the Optimization of Atomic Layer Deposition},
	type = {doi: 10.1021/acsami.1c00649},
	url = {https://doi.org/10.1021/acsami.1c00649},
	volume = {13},
	year = {2021},
	year1 = {2021}}

@article{YanguasGil2025,
	author = {Yanguas-Gil, Angel and Dearing, Matthew T. and Elam, Jeffrey W. and Jones, Jessica C. and Kim, Sungjoon and Mohammad, Adnan and Thang Nguyen, Chi and Sengupta, Bratin},
	doi = {10.1116/6.0004319},
	issn = {0734-2101},
	journal = {Journal of Vacuum Science \& Technology A},
	month = {04},
	number = {3},
	pages = {032406},
	title = {Benchmarking large language models for materials synthesis: The case of atomic layer deposition},
	url = {https://doi.org/10.1116/6.0004319},
	volume = {43},
	year = {2025}}

@article{Klepper2007,
	author = {K.B. Klepper and O. Nilsen and H. Fjellv{\aa}g},
	doi = {10.1016/j.tsf.2007.03.182},
	issn = {0040-6090},
	journal = {Thin Solid Films},
	number = {20},
	pages = {7772-7781},
	title = {Growth of thin films of {Co3O4} by atomic layer deposition},
	url = {https://www.sciencedirect.com/science/article/pii/S0040609007005196},
	volume = {515},
	year = {2007}}

@article{Tero2009,
	author = {Pilvi, Tero and Puukilainen, Esa and Munnik, Frans and Leskel{\"a}, Markku and Ritala, Mikko},
	doi = {10.1002/cvde.200806721},
	eprint = {https://onlinelibrary.wiley.com/doi/pdf/10.1002/cvde.200806721},
	journal = {Chemical Vapor Deposition},
	number = {1-3},
	pages = {27-32},
	title = {{ALD} of {YF3} Thin Films from {TiF4} and {Y(thd)3} Precursors},
	url = {https://onlinelibrary.wiley.com/doi/abs/10.1002/cvde.200806721},
	volume = {15},
	year = {2009}}

@article{Comstock2012,
	author = {Comstock, David J. and Elam, Jeffrey W.},
	date = {2012/11/13},
	doi = {10.1021/cm300712x},
	isbn = {0897-4756},
	journal = {Chemistry of Materials},
	journal1 = {Chemistry of Materials},
	journal2 = {Chem. Mater.},
	month = {11},
	number = {21},
	pages = {4011--4018},
	publisher = {American Chemical Society},
	title = {Atomic Layer Deposition of {Ga2O3} Films Using Trimethylgallium and Ozone},
	type = {doi: 10.1021/cm300712x},
	url = {https://doi.org/10.1021/cm300712x},
	volume = {24},
	year = {2012},
	year1 = {2012}}

@article{Devika2020,
	author = {Choudhury, Devika and Mandia, David J. and Langeslay, Ryan R. and Yanguas-Gil, Angel and Letourneau, Steven and Sattelberger, Alfred P. and Balasubramanium, Mahalingam and Mane, Anil U. and Delferro, Massimiliano and Elam, Jeffrey W.},
	doi = {10.1116/6.0000053},
	issn = {0734-2101},
	journal = {Journal of Vacuum Science \& Technology A},
	month = {06},
	number = {4},
	pages = {042407},
	title = {Atomic layer deposition of {HfO2} films using carbon-free tetrakis(tetrahydroborato)hafnium and water},
	volume = {38},
	year = {2020}}

@article{Aaltonen2005,
	author = {Aaltonen, Titta and Ritala, Mikko and Leskel{\"a}, Markku},
	doi = {10.1149/1.1940507},
	journal = {Electrochemical and Solid-State Letters},
	month = {jun},
	number = {8},
	pages = {C99},
	publisher = {The Electrochemical Society, Inc.},
	title = {{ALD} of Rhodium Thin Films from {Rh}(acac)3 and Oxygen},
	url = {https://doi.org/10.1149/1.1940507},
	volume = {8},
	year = {2005}}

\end{document}